\documentclass[12pt,preprint]{aastex}

\newcommand{\lsim}{\raise0.3ex\hbox{$<$}\kern-0.75em{\lower0.65ex\hbox{$\sim$}}}
\newcommand{\gsim}{\raise0.3ex\hbox{$>$}\kern-0.75em{\lower0.65ex\hbox{$\sim$}}}
\newcommand{\propsim}{\raise0.3ex\hbox{$\propto$}\kern-0.75em{\lower0.65ex\hbox{$\sim$}}}

\begin{document}
    
\title{Circular polarisation from inhomogeneous synchrotron sources}

\author{C.-I. Bj\"ornsson\altaffilmark{1}}
\altaffiltext{1}{Department of Astronomy, AlbaNova University Center, Stockholm University, SE--106~91 Stockholm, Sweden.}
\email{bjornsson@astro.su.se}

\begin{abstract}
Inhomogeneities can influence the polarisation emerging from a synchrotron source. However, it is shown that the frequency distribution of circular polarisation is only marginally affected, although its magnitude may change substantially. This is used to argue that the observed properties of compact radio sources imply a radiating plasma in which the characteristic waves are nearly circular. As a result, restrictions can be put on the low energy part of the energy distribution of the relativistic electrons as well as the presence of electron-positron pairs. It is emphasised that this constrains theoretical modelling of the acceleration process for the relativistic electrons; for example, some of the currently popular scenarios seem to need modifications to become consistent with observations.
\end{abstract}


\section{Introduction}
It is commonly believed that the launching of jets in AGNs is driven by magnetic fields \citep{b/z77,b/p82}. At some distance down-stream of the jet, some fraction of the Poynting flux needs to be converted into kinetic energy and relativistic particles in order to give rise to the observed radiation. The understanding of the processes by which this occurs is still rather limited. 

Several mechanisms have been suggested for the acceleration of the particles; for example, diffusive shock acceleration, second order Fermi acceleration in a turbulent medium \citep[e.g.,][]{zhd18,bla19} and magnetic reconnection \citep[e.g.,][]{r/l92,gia09}. Any of these mechanisms can give rise to a particle energy distribution, which, at the high end, is consistent with observations. On the other hand, their low energy part is expected to differ, since it reflects the injection of particles into the acceleration process. Unfortunately, this part is normally hidden from view by optical depth effects. 

Another related issue concerns the composition of the plasma, in particular, the presence of electron-positron pairs. As discussed by \cite{sik97} and \cite{s/k00}, a pure electron-positron plasma is likely to overproduce the X-ray emission through the bulk comptonization of low energy photons in the most luminous sources. At the same time, the kinetic energy of a jet dominated by electron-protons is often deduced to exceed that released through the accretion process \citep{ghi14,maj16}. In order to be consistent with both of these constraints, it has been argued that the plasma needs to contain, roughly, ten electron-positron pairs per proton \citep{ghi10,maj16}. Furthermore, these two issues are connected, since the relative number of electron-positron pairs is expected to influence both particle heating and the efficiency of the acceleration process \citep{pet19}.

Although neither the low energy electrons nor the presence of electron-positron pairs can be observed directly, they can significantly affect the propagation of polarised light through a medium; this is particularly true for circular polarisation. The potential importance of circular polarisation in compact radio sources was realised early on \citep[e.g.,][]{pac73}. However, the low observed value limited its role as a plasma diagnostic. To some extent, this has now changed with the advent of more sensitive observations \citep{mac00, ray00}, a larger frequency range \citep{osu13, agu18}, and larger spatial resolution through VLBI \citep{h/w04,hom09}. This was used in \cite{bjo19} to argue that the observed properties of the circular polarisation indicate that the characteristic waves are nearly circularly polarised rather than nearly linearly in the emitting plasma. This puts quite strong constraints on the combined properties of the electron energy distribution and the presence of electron-positron pairs.

The conclusion in \cite{bjo19} relied, mainly, on the frequency distribution of the circular polarisation from a homogeneous source. The aim of the present paper is to determine to what extent inhomogeneities may alter the polarisation properties emerging from a synchrotron source. The transport of polarised light is expressed as a coupling between the propagating characteristic waves \citep{for42}, which is in contrast to the standard way of using Stokes parameters. This allows for a more transparent discussion of the physical effects and, in particular, the solution of the transport equations can be expressed in terms of the polarisation properties of the characteristic waves. 

The paper is structured as follows: The coupling of characteristic waves is introduced in Section\,\ref{sect2}. It is shown that the ensuing transport equation can be obtained in a simpler and more concise way than is usually done. Its solution is discussed in Section\,\ref{sect3}. A constant coupling approximation is used to bring forth the general properties resulting from inhomogeneities. The main conclusion is that they can significantly affect the emerging circular polarisation. However, this applies mainly to its magnitude, the frequency dependence is only marginally affected. In Section\,\ref{sect4}, this approximation is contrasted with another one in which a given change in plasma properties is modelled as occurring instantaneously. Although the resulting polarisation can be quite different, again, the frequency distribution of the circular polarisation is not expected to be seriously affected. A discussion of the results follows in Section\,\ref{sect5}. It is emphasised that, independent of the presence of inhomogeneities, the frequency distribution of the circular polarisation should be a good discriminator between various plasma properties. The main points of the paper are summarised in Section\,\ref{sect6}.

\section{A more concise derivation of the transport equation for polarised light in an inhomogeneous medium} \label{sect2}
\begin{figure}
\epsscale{0.80}
\plotone{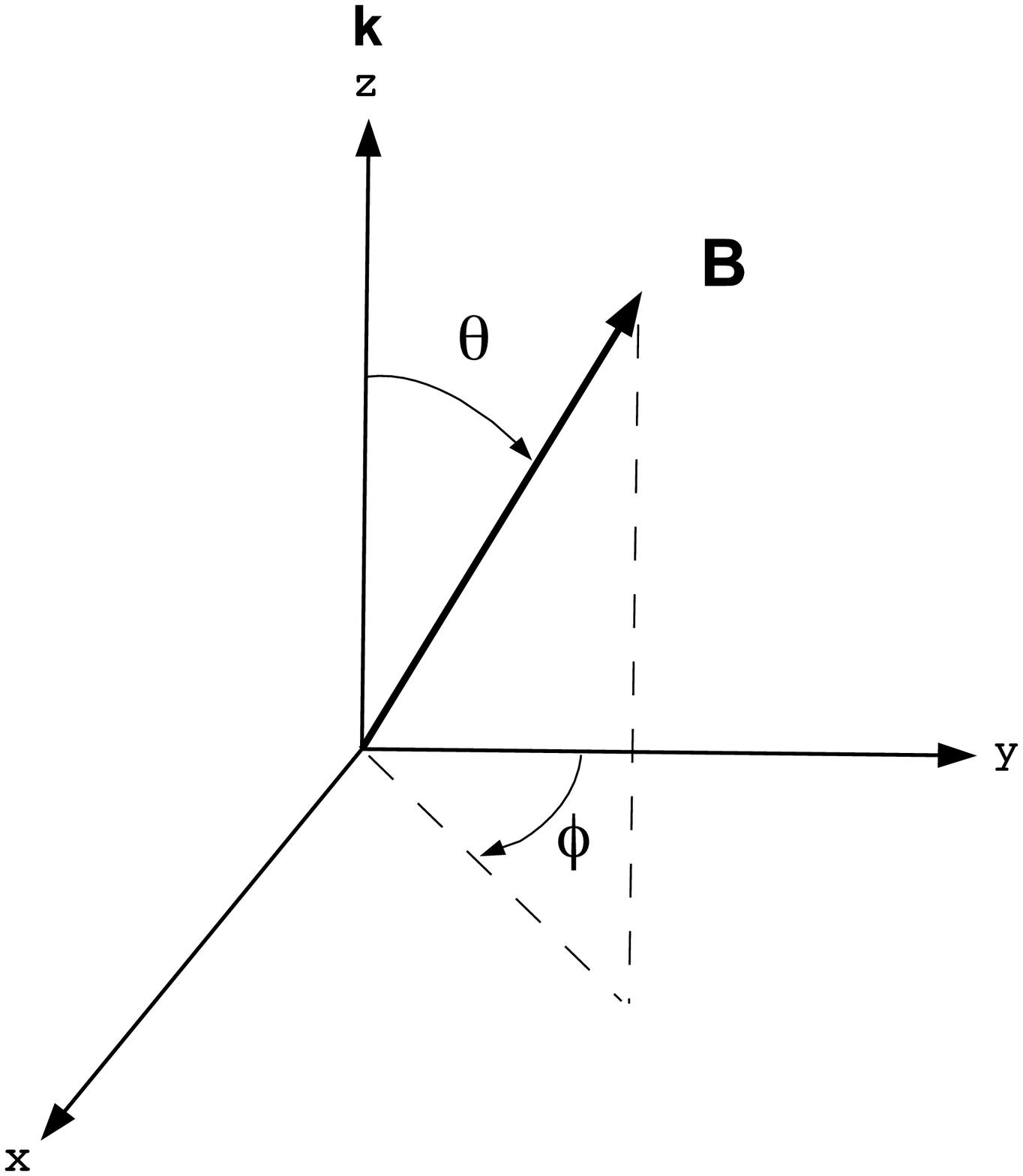}
\caption{The coordinate system used for the transfer equation. The ray propagates along the z-axis and the magnetic field direction is specified by the polar-angle $\theta$ and azimuthal-angle $\phi$.
\label{fig1}} 
\end{figure}

Locally, the interaction between a propagating electromagnetic field and the plasma can be described by $J_{\rm l} = \sigma_{\rm l,m} E_{\rm m}$. Here, $J_{\rm l}$ is the current, $E_{\rm m}$ is the electric field, and $\sigma_{\rm l,m}$ is the dielectric tensor. The indices (l,m) run over the three spatial coordinates, i.e., (l,m = x,y,z) and a repeated index indicates summation. The notation in this paper follows that in \cite{bjo19} \citep[see also][]{j/o77}. The amplitude of the electric field in the direction of its propagation (the z-direction, see Figure\,\ref{fig1}) is only a fraction $|\sigma_{\rm l,m}/\nu| \sim \kappa \lambda$ of that in the perpendicular direction, where $\nu$ and $\lambda$ are the frequency and wavelength of the electromagnetic wave, respectively, while $\kappa$ is the absorptivity of the medium. Since this is usually a very small number, the propagation of the electromagnetic field can be approximated by a second order partial differential equation for a two-dimensional plane wave (i.e., l,m = x,y). Furthermore, as discussed in \cite{bjo19}, without any additional approximations, this wave-equation can be further reduced to a first order ordinary differential equation
\begin{equation}
	\frac{\rm d}{{\rm d}s}\, E_{\rm l} = -\frac{2\pi}{c}\sigma_{\rm l,m}E_{\rm m},
	\label{eq1}
\end{equation}
where $s$ is the distance along a ray path. 

The local properties of the plasma can be used to define two characteristic waves ($1\&2$)
\begin{equation}
	J_{\rm l}^{\rm 1,2} = \eta^{\rm 1,2}E_{\rm l}^{\rm 1,2},
	\label{eq2}
\end{equation}
where $ \eta^{\rm 1,2}$ are the two eigenvalues obtained by diagonalising $\sigma_{\rm l,m}$. Furthermore, $J_{\rm l} = J_{\rm l}^{\rm 1} +J_{\rm l}^{\rm 2}$ and $E_{\rm l} = E_{\rm l}^{\rm 1} + E_{\rm l}^{\rm 2}$. The plasma properties can be described by $\Upsilon_{\rm V} = \hat{\xi}_{\rm V} + i \xi_{\rm V}$, which accounts for the circular birefringence and absorption, and the corresponding linear quantity $\Upsilon_{\rm L} = \hat{\xi}_{\rm U} + i \xi_{\rm U}$ \citep[see][]{bjo19}. The eigenvalues can then be expressed as 
\begin{equation}
	\eta^{1,2} = \frac{c\kappa}{4\pi}\left(1 \mp i\sqrt{\Upsilon_{\rm V}^2 +\Upsilon_{\rm L}^2}\right).
	\label{eq3}
\end{equation}
Furthermore,
\begin{equation}
	K^{1,2} =  \frac{\pm\sqrt{1-\rho^2} - \sin(2\varphi)}{\rho +\cos(2\varphi)},
	\label{eq4}
\end{equation}
where $K^{1,2} \equiv E_{\rm y}^{1,2}/E_{\rm x}^{1,2}$ are the polarisation of the two characteristic waves \citep{j/o77b} and $\rho \equiv i \Upsilon_{\rm V}/\Upsilon_{\rm L}$. Here, the azimuthal angle vary along the ray path as $\phi = -\pi/4 + \varphi$, with $\varphi = 0$ for $s=0$ (see Figure\,\ref{fig1}).

The standard transport equation expressed in terms of the Stokes parameters can be obtained directly from Equation (\ref{eq1}). There are two aspects of this equation that should be noted; namely, (1) it is valid also for inhomogeneous media and (2) the notion of characteristic waves does not enter in its formulation. However, a not so attractive property is its low physical transparency. It was \cite{for42}, who first suggested the use of characteristic waves to elucidate the effects that inhomogeneities have on the polarisation properties of a propagating electromagnetic wave. The reason for this approach is that for a homogeneous medium, the characteristic waves propagate independently and the solution to Equation (\ref{eq1}) can be written $\mathbf{E} \equiv (E_{\rm x}, E_{\rm y}) = (E_{\rm x}^{\rm 1} +E_{\rm x}^{\rm 2}, K^1E_{\rm x}^{\rm 1} + K^2 E_{\rm x}^{\rm 2})$, where
\begin{equation}
	\mathbf{E}^{1,2} = \mathbf{E}^{1,2}_{\rm o} \exp\left(-\frac{2\pi}{c}\eta^{1,2} s \right).
	\label{eq5}
\end{equation}
In this formulation, the effects of inhomogeneities manifest themselves as a coupling between the two characteristic waves. This idea was further developed by \cite{coh60}. The WKB-approximation to the original wave equation was discussed by \cite{gin61}. Physically, in this approximation, the polarisation of the propagating characteristic waves adjusts to their local values; hence, no coupling between them occurs. However, the conditions for its applicability can easily be violated. Instead, it has been used as an "ansatz" to derive the coupling between the characteristic waves \citep[see e.g.,][]{j/o77b}.

The derivation usually takes as its starting point the wave-equation, i.e.,  a second order differential equation. This leads to a long and rather tedious calculation, where, in the end, only terms of lowest order in the small quantity $\kappa \lambda$ are retained. However, as discussed above, without loss of accuracy, one may instead start with Equation (\ref{eq1}). Since this is a first order ordinary differential equation, a substantially shorter derivation should result. Its solution for a homogeneous medium (Equation \ref{eq5}) contains two constant $\mathbf{E}_{\rm o}^{1,2}$. A suitable "ansatz" for the general solution is then 
\begin{equation}
	E_{\rm x}^{1,2} = \bar{E}^{1,2} \exp\left(-\frac{2\pi}{c}\int_0^{s} \eta^{1,2} {\rm d}
	\hat{s} \right),
	\label{eq6}
\end{equation}
where the spatial variations of $\eta^{1,2}$ (and $K^{1,2}$) now imply that also $\bar{E}^{1,2}$  vary with distance along a ray path.

Equation (\ref{eq1}) can then be rewritten as
\begin{equation}
	\frac{\rm d}{{\rm d}s} (E_{\rm x}^1 + E_{\rm x}^2) = -\frac{2 \pi}{c}(\eta^1 E_{\rm x}^1 +
	\eta^2 E_{\rm x}^2),
	\label{eq7}
\end{equation}
for the x-component and similarly for the y-component
\begin{equation}
	\frac{\rm d}{{\rm d}s} (K^1 E_{\rm x}^1 + K^2 E_{\rm x}^2) = -\frac{2 \pi}{c}(\eta^1 K^1 
	E_{\rm x}^1 +\eta^2 K^2 E_{\rm x}^2).
	\label{eq8}
\end{equation}
With the use of Equation (\ref{eq6}) one then finds
\begin{eqnarray}
	\frac{\rm d}{{\rm d}s} (\bar{E}^1) +  \exp \left(-\int_0^{s} \Delta k {\rm d}\hat{s}\right)
	 \frac{\rm d}{{\rm d}s} (\bar{E}^2) & = & 0\nonumber\\
	\frac{\rm d}{{\rm d}s} (K^1 \bar{E}^1) +  \exp\left(-\int_0^{s} \Delta k {\rm d}
	\hat{s}\right) \frac{\rm d}{{\rm d}s} (K^2\bar{E}^2) & = & 0,
	\label{eq9}
\end{eqnarray}
where $\Delta k \equiv -(2\pi / c)(\eta^1-\eta^2)$ is the phase difference between the two characteristic waves.

Substituting the expression for $(\rm d/{\rm d}s) \bar{E}^1$ from the first part into the second part of Equation (\ref{eq9}) yields
\begin{equation}
	\frac{\rm d}{{\rm d}s} \bar{E}^2 + \frac{{\rm d}K^2}{{\rm d}s}\frac{\bar{E}^2}{K^2-K^1}
	 = - \frac{{\rm d}K^1}{{\rm d}s}\frac{\bar{E}^1}{K^2-K^1}
	\exp \left(\int_0^{s} \Delta k {\rm d}\hat{s}\right),
	\label{eq10}
\end{equation}
which can be rewritten as
\begin{equation}
	\frac{\rm d}{{\rm d}s}\left[\bar{E}^2 \exp \int_0^{s}\frac{{\rm d}K^2}{{\rm d}\hat{s}}
	\frac{{\rm d}\hat{s}}{(K^2 - K^1)}\right] = -\frac{{\rm d}K^1}{{\rm d}s}\frac{\bar{E}^1}{K^2 - K^1}
	\exp \int_0^{s}\left(\frac{{\rm d}K^2}{{\rm d}\hat{s}}
	 \frac{1}{(K^2 - K^1)} +\Delta k \right){\rm d}\hat{s}. 
	\label{eq11}
\end{equation}
The complementary equation is obtained by instead substituting $(\rm d/{\rm d}s) \bar{E}^2$ from the first part into the second part of Equation (\ref{eq9}). It is seen that the corresponding equation can be obtained directly from Equation (\ref{eq11}) by interchanging 1 \& 2 (i.e., $1 \leftrightarrow 2$) and letting $\Delta k \rightarrow -\Delta k$.

The coupling between the characteristic waves is normally expressed in terms of the amplitudes of the WKB-approximation, which are given by
\begin{equation}
	E_{\rm WKB}^{1,2} = \bar{E}^{1,2}\exp \int_0^{s}\frac{{\rm d}K^{1,2}}{{\rm d}\hat{s}}
	 \frac{{\rm d}\hat{s}}{(K^{1,2} - K^{2,1})}.
	 \label{eq12}
\end{equation}
The standard formulation of the propagation of electromagnetic waves in an inhomogeneous medium is then obtained from Equation (\ref{eq12}) and its complement as
\begin{eqnarray}
	\frac{{\rm d} E^1_{\rm WKB}}{{\rm d} s} & = & \Psi^2 E^2_{\rm WKB} \exp\left(-\int_0^{s}
	\Delta k {\rm d}\hat{s}\right)\nonumber\\
	\frac{{\rm d} E^2_{\rm WKB}}{{\rm d} s} & = & \Psi^1 E^1_{\rm WKB} \exp\left(\int_0^{s}
	\Delta k {\rm d}\hat{s}\right),
	\label{eq13}
\end{eqnarray}
where
\begin{eqnarray}
	\Psi^1 & = & \frac{{\rm d} K^1}{{\rm d} s}\frac{1}{K^1 - K^2} \exp \int_0^{s}
	\left(\frac{{\rm d} K^1}{{\rm d} \hat{s}} + \frac{{\rm d} K^2}{{\rm d} \hat{s}}\right) \frac{{\rm d}
	\hat{s}}{K^2 - K^1}\nonumber\\
	\Psi^2 & = & \frac{{\rm d} K^2}{{\rm d} s}\frac{1}{K^2 - K^1} \exp \int_0^{s}
	\left(\frac{{\rm d} K^1}{{\rm d} \hat{s}} + \frac{{\rm d} K^2}{{\rm d} \hat{s}}\right) \frac{{\rm d}
	\hat{s}}{K^1 - K^2}
	\label{eq14}
\end{eqnarray}
are the two coupling parameters.

The relative ease with which Equations (\ref{eq13}) and (\ref{eq14}) are derived does not only
depend on the starting point. It may also be noticed that no higher order terms in $\kappa \lambda$ occur in the derivation, i.e., all higher order terms disappear when starting from Equation (\ref{eq1}) rather than the wave-equation. Furthermore, the calculations are made less cumbersome by a functionally simpler "ansatz" (Equation (\ref{eq6}) instead of the WKB-approximation).
  
Although Equation (\ref{eq13}) explicitly shows how the inhomogeneities couple the characteristic waves, the underlying physics can be made more transparent by introducing
\begin{equation}
	\tilde{E}^{1,2} = E^{1,2}_{\rm WKB} \exp \left(\pm \int_0^{s}\frac{\Delta k}{2} {\rm d}\hat{s}\right)
	\label{eq15}
\end{equation}
so that the Equation (\ref{eq13}) can be written
\begin{eqnarray}
	\frac{{\rm d}\tilde{E}^1}{{\rm d}s} - \frac{\Delta k}{2}\tilde{E}^1 & = & \Psi^2 \tilde{E}^2
	\nonumber\\
	\frac{{\rm d}\tilde{E}^2}{{\rm d}s} + \frac{\Delta k}{2}\tilde{E}^2 & = & \Psi^1 \tilde{E}^1.
	\label{eq16}
\end{eqnarray}
The solution to Equation (\ref{eq1}) is then
\begin{eqnarray}
	E_{\rm x}^1 = \tilde{E}^1\exp \int_0^{s}\frac{{\rm d}K^1}{{\rm d}\hat{s}}
	 \frac{{\rm d}\hat{s}}{(K^2 - K^1)} \exp\left(-\int_0^{s}\frac{\kappa}{2}{\rm d}\hat{s}\right)
	 \nonumber\\		
	 E_{\rm x}^2 = \tilde{E}^2\exp \int_0^{s}\frac{{\rm d}K^2}{{\rm d}\hat{s}}
	 \frac{{\rm d}\hat{s}}{(K^1 - K^2)} \exp\left(-\int_0^{s}\frac{\kappa}{2}{\rm d}\hat{s}\right).
	 \label{eq17}
\end{eqnarray}
It is directly seen from Equation (\ref{eq16}) that the coupling between the characteristic waves is determined by the ratio $|\Psi^{1,2}/\Delta k|$. The limit $|\Psi^{1,2}|\gg |\Delta k|$ implies that the coupling is so strong that the plasma properties vary more rapidly along a ray path than does the relative phase between the characteristic waves. Effectively, then, the plasma is isotropic, and the polarisation stays roughly constant. Apart from this limit, the emerging polarisation is due to an interplay between the adiabatically changing properties of the characteristic waves (i.e., the WKB-approximation) and their coupling. In order to highlight the effects of the inhomogeneities, the result of this interplay will be presented as deviations from that of a homogeneous medium. 

The interaction between these two independent effects may lead to the conclusion that the polarisation is only rarely seriously affected by inhomogeneities. However, it is important to note that the relative change of a small quantity can be significant over a much wider range of conditions. For synchrotron emission, the intrinsic degree of circular polarisation is much smaller than that of the linear polarisation. Hence, inhomogeneities may influence the circular polarisation, while having a negligible effect on the linear polarisation. This is analogous to the homogeneous case, where the degree of ellipticity of the characteristic waves causes the conversion of linear to circular polarisation; for example, only a small deviation from either circularly or linearly polarised characteristic wave can result in circular polarisation significantly different from the intrinsic one \cite[e.g.,][]{bjo19}.

\section{Solution to the transport equation}\label{sect3}

Numerical solutions to Equation (\ref{eq13}) were discussed in \cite{bjo90}. Since the aim of the present paper is to bring forth the underlying physics governing the effects of inhomogeneities, a different approach is followed below. Two limiting situations will be considered. The first assumes $\phi = constant$. This makes it possible to choose $K^1 = - K^2$ (see Equation \ref{eq4}), which implies $\Psi^1 = \Psi^2 = (1/2)({\rm d} \ln K^{1,2} / {\rm d} s)$ (Equation \ref{eq14}). In the general case, $\phi$ also varies. When the transport effects are dominated by variations in $\phi$, it is shown that they are well described by $\Psi^1 = - \Psi^2$ over a limited range in $\phi$.

\subsection{Magnetic field with a constant azimuthal angle}\label{sect3a}
This case was discussed in some detail in \cite{bjo90}. The focus in this section is therefore limited to  its generic properties and how  inhomogeneities may modify the conclusions drawn from the homogeneous solution. In order to simplify the notation, $K\equiv K^2$ and $\Psi \equiv \Psi^1 = \Psi^2$ will be used. The solution to Equation (\ref{eq1}) can then be written
\begin{eqnarray}
	E_{\rm x} =E_{\rm x}^1 + E_{\rm x}^2 & = &  (\tilde{E}^1 + \tilde{E}^2)\sqrt \frac{K_{\rm o}}{K} 	
	\exp\left(-\int_0^{s}\frac{\kappa}{2}{\rm d}\hat{s}\right)\nonumber\\
	E_{\rm y} =E_{\rm y}^1 + E_{\rm y}^2 & = &  -(\tilde{E}^1 - \tilde{E}^2)\sqrt {K_{\rm o}K}	
	\exp\left(-\int_0^{s}\frac{\kappa}{2}{\rm d}\hat{s}\right),
	\label{eq18}
\end{eqnarray}
where $K_{\rm o}$ is the value of $K$ at $s=0$.
Introducing, $X \equiv \tilde{E}^1 + \tilde{E}^2$ and $Y \equiv \tilde{E}^1 - \tilde{E}^2$ together with $\alpha = \Delta k / 2 \Psi$ and ${\rm d}\chi = \Psi {\rm d}s$, Equation (\ref{eq16}) can be written
\begin{eqnarray}
	\frac{{\rm d}X}{{\rm d}\chi} & = & X + \alpha Y
	\nonumber\\
	\frac{{\rm d}Y}{{\rm d}\chi} & = & \alpha X -Y.
	\label{eq19}
\end{eqnarray}

With $\alpha = constant$, Equation (\ref{eq19}) is analogous to the equation for propagation in a homogeneous medium and can be solved in a similar way. This is done in Appendix A, where it is shown that
\begin{eqnarray}
	E_{\rm x} & = & \sqrt {\frac{I_{\rm o}}{2(1+q_{\rm o})}}\left((1+q_{\rm o} - \alpha\sigma_{\rm o})\frac{\sinh(\beta \chi)}{\beta}+ 
	(1+q_{\rm o})\cosh(\beta \chi)\right)\exp(-\chi-\tau/2)\nonumber\\
	E_{\rm y} & = & -K_{\rm o}\sqrt {\frac{I_{\rm o}}{2(1+q_{\rm o})}}\left((\alpha(1+q_{\rm o}) +\sigma_{\rm o})\frac{\sinh(\beta \chi)}{\beta}
	 -  \sigma_{\rm o}\cosh(\beta \chi)\right) \exp(\chi -\tau/2).\nonumber\\
	\label{eq20}
\end{eqnarray}
The emitted emission at $s=0$ is assumed to be $100 \%$ polarised and described by the Stokes parameters $I_{\rm o},Q_{\rm o},U_{\rm o}$, and $V_{\rm o}$. Here,  $q_{\rm o}=Q_{\rm o}/I_{\rm o}$, $u_{\rm o}=U_{\rm o}/I_{\rm o}$, $v_{\rm o}=V_{\rm o}/I_{\rm o}$, so that $q_{\rm o}^2 + u_{\rm o}^2 + v_{\rm o}^2 = 1$. Furthermore, $\beta \equiv \sqrt {1+\alpha^2}$, $\sigma_{\rm o} \equiv (u_{\rm o} - iv_{\rm o})/K_{\rm o}$, $\chi = \int_o^s \Psi {\rm d}\hat{s} = (\ln {K/K_{\rm o}})/2$ and $\tau = \int_0^s\kappa {\rm d}{\hat s}$ is the optical depth along the ray path. With $\alpha$ assumed to be a constant, it can be expressed as $\alpha = \delta k \tau/2\chi$, where $ \delta k \tau \equiv \int_0^s \Delta k {\rm d}\hat{s} = \int_0^\tau (\Delta k/\kappa){\rm d}\hat{\tau}$. It is also shown in Appendix A, that the approximation $\alpha = constant$ should be a good one as long as $(1+\alpha^2)^{-1} ({\rm d}\ln \alpha /{\rm d} \chi) = 2(1+\alpha^2)^{-1} ({\rm d}\alpha /\Delta k{\rm d}s) \ll 1$.

With
\begin{equation}
	\Psi = \frac{1}{2} \frac{{\rm d} \ln K}{{\rm d}s} = -\frac{1}{2(1-\rho^2)}\frac{{\rm d} \rho}
	{{\rm d} s}, 
	\label{eq21}
\end{equation}
it is seen that a necessary condition for inhomogeneities to produce large values of $\chi$ (i.e., $|\chi| \sim 1$) is that $|\rho| \sim 1$. As discussed in Appendix C, this, in turn, implies large values for the circular polarisation in a homogeneous source. However, the observed values in compact radio sources is usually quite small ($\lsim 1\%$). It is argued in Section\,\ref{sect5b} that this low value is unlikely to be the result of large scale cancellation due to an inhomogeneous source structure. If so, one may conclude that $|\chi|$ is substantially smaller than unity and, hence, that either $|K_{\rm i}| \ll 1$ (i.e., $|\rho| \ll 1$, nearly linear characteristic waves) or $|K_{\rm r}| \ll 1$ (i.e., $|\rho| \gg 1$, nearly circular characteristic waves). 

In order to estimate the importance of inhomogeneities, it is instructive to expand Equation (\ref{eq20}) to lowest order in $|\beta \chi|$, and this yields
\begin{eqnarray}
	E_{\rm x} & = & \sqrt {\frac{I_{\rm o}}{2(1+q_{\rm o})}}\left((1+q_{\rm o} -\sigma_{\rm o}\frac{\delta k \tau}{2}(1-\chi)\right)\exp(-\tau/2)
	\nonumber\\
	E_{\rm y} & = & K_{\rm o}\sqrt {\frac{I_{\rm o}}{2(1+q_{\rm o})}}\left(\sigma_{\rm o} - (1+q_{\rm o})\frac{\delta k \tau}{2}
	 (1+\chi)\right) \exp( -\tau/2).\nonumber\\
	\label{eq21a}
\end{eqnarray}
With $|\chi|\ll 1$, the limit $|\beta \chi| \sim 1$ corresponds to $|\alpha| \sim |\chi|^{-1} \gg1$ and $|\delta k| \tau \sim 1$. It should be noted that the first order terms in $\chi$ (i.e., $u_{\rm o} \chi$) have cancelled. Equation (\ref{eq21a}) is a valid approximation for all values of $|\alpha|$, since $|\alpha|<1$ implies $\beta \approx 1$ and $|\alpha|> 1$ leads to $\chi \delta k \tau/2 = \alpha \chi^2$ so that $|\chi|^2$-terms can be neglected in this limit. 

Since $2E_{\rm x}E_{\rm y}^{*} \equiv U + iV$, where (*) denotes complex conjugate, it is shown in Appendix A that Equation (\ref{eq21a}) leads to
\begin{equation}
	V = I_{\rm o}[v_{\rm o} - \xi_{\rm V}\tau - q_{\rm o}\hat{\xi}_{\rm U}\tau - \hat{\xi}_{\rm U}\chi_{\rm r}\tau + \xi_{\rm U}\chi_{\rm i}\tau
	- q_{\rm o} (\hat{\xi}_{\rm V}\chi_{\rm i} + \xi_{\rm V}\chi_{\rm r})\tau],
	\label{eq21b}
\end{equation}
which gives the circular polarisation in the limit $|\delta k|\tau < 1$. The subscripts "r" and "i" are used to denote the real and imaginary parts, respectively, of a quantity. Furthermore, the various $\xi$-parameters in Equation (\ref{eq21b}) are now quantities integrated over the ray path, so that $\xi \tau$ stands for $\int_0^\tau \xi{\rm d}\hat{\tau}$. With this redefinition of the $\xi$-parameters, the polarisation can be considered to consist of one "homogeneous" part and one due to the inhomogeneities (i.e., $\chi$). For synchrotron radiation, $|\xi_{\rm V}| \ll 1$ and $q_{\rm o} = 0$. Although the homogeneous terms contributing to the circular polarisation are $\sim |\delta k|$, it is seen from Equation (\ref{eq21b}) that cancellation occurs, and their net result is only $|\xi_{\rm V}|$ \citep[see][for a more detailed discussion]{bjo19}. 

A similar cancellation also takes place for the second order inhomogeneous terms. For $|\rho| \ll 1$, the inhomogeneous terms are $\sim |\chi|$. Writing $K = -1 + \Delta K$, with $|\Delta K| \ll 1$, one finds $\chi = (\Delta K_{\rm o} - \Delta K)/2$. It was shown in \cite{bjo19} that locally $-\hat{\xi}_{\rm U} \Delta K_{\rm r} + \xi_{\rm U}\Delta K_{\rm i} = \xi_{\rm V}$. Hence, the inhomogeneous terms in Equation (\ref{eq21b}) correspond, roughly, to the variations of $\xi_{\rm V}$ along the ray path (i.e., a very small number). Furthermore, in the limit $|\rho|\gg 1$, $|\hat{\xi}_{\rm V}| \gg 1$ and dominates all the other $\xi$-parameters. Since Equation (\ref{eq21b}) is valid for $\tau \lsim |\hat{\xi}_{\rm V}|^{-1}$, one finds,  in this case, that the contribution from the inhomogeneities are $\sim |\chi|/|\hat{\xi}_{\rm V}|$. This shows that for frequencies such that $\tau \lsim |\delta k|^{-1}$, transport effects are likely to only marginally affect the emerging circular polarisation (i.e., its value is given by $\sim v_{\rm o}$). This is true whether or not the plasma is inhomogeneous. As shown below, these cancellations do not occur for $|\delta k|\tau \gsim 1$; hence, one expects a rapid increase in the circular polarisation for frequencies corresponding to the range where $\tau \sim |\delta k|^{-1}$. 

Another useful expression for the circular polarisation can be obtained from Equation (\ref{eq20}) for large values of $|\alpha|$. Expansion to first order in $|\alpha|^{-1}$ yields (see Appendix A)
\begin{eqnarray}
	U + iV & = & I_{\rm o} \exp(-\tau -2i\chi_{\rm i}) \times 
	\nonumber\\ 
	& &  \left[u_{\rm hom} + iv_{\rm hom} +\frac{K_{\rm o}^{*}}
	{|\alpha|^2}\left \{\frac{\alpha(1-q_{\rm o})}{|K_{\rm o}|^2} - \alpha^{*}(1+q_{\rm o})\right\} 
	|\sinh (\delta k \tau/2)|^2\right.
	\nonumber\\
	& & + \left. \frac{2i(u_{\rm o}+iv_{\rm o})}{|\alpha|^2}\{\alpha^{*}
	\sinh (\delta k \tau/2) \cosh ^{*} (\delta k \tau/2)\}_{\rm i}\right].
	\label{eq22}
\end{eqnarray}
Here, $u_{\rm hom}+iv_{\rm hom}$ is the part corresponding to the homogeneous case with an optical depth $\tau$. The different effects of $\chi$ (the WKB-approximation) and $\alpha$ (the coupling between the characteristic waves) are clearly seen. While $\chi$ gives rise to circular polarisation through a process similar to conversion of linear polarisation in a homogeneous plasma, $\alpha$ accounts for the circular polarisation induced by the interaction with the local medium. One may notice two things from Equation (\ref{eq22}): (1) It describes the emerging polarisation of a light ray with optical depth $\tau$ and initial values $I_{\rm o}, Q_{\rm o}, U_{\rm o}, V_{\rm o}$, and $K_{\rm o}$. The polarisation of the total radiation is obtained by integrating along the line of sight, i.e., over $\tau$ and taking the variations of the initial values into account. (ii) It is a linear function of the initial Stokes parameters. Since Stokes parameters are additive, Equation (\ref{eq22}) is valid also for partially polarised light rays. This is generally true \citep{bjo88}.

The range of validity of Equation (\ref{eq22}) overlaps that of Equation (\ref{eq21a}) ($1<|\alpha|<|\chi|^{-1}$); hence, the latter can be obtained by expanding the former for $|\delta k|\tau < 1$. In the opposite limit (i.e., $|\delta k|\tau > 1$), the terms $\propto \cos(\delta k_{\rm i} \tau)$ and $\propto \sin(\delta k_{\rm i} \tau)$  are unlikely to be important, since integration along the line of sight tends to cancel out their contributions. With this simplification, Equation (\ref{eq22}) can be written to first order in $|\chi|$
\begin{eqnarray}
	U + iV  &=& I_{\rm o} \exp(-\tau)\times\nonumber\\
	&&\left[u_{\rm hom} + iv_{\rm hom} -2i\chi_{\rm i}u_{\rm hom} + \frac{i\alpha_{\rm i}}{|\alpha|^2}\{K_{\rm o}^{*}\cosh(\delta k_{\rm r} \tau) - 
	u_{\rm o}\sinh(\delta k_{\rm r} \tau)\}\right.\nonumber\\ &&\hspace*{2.5cm}-\left.\frac{q_{\rm o}K_{\rm o}^{*}\alpha_{\rm r}}{|\alpha|^2}\cosh(\delta k_{\rm r} \tau)\right],
	\label{eq23}
\end{eqnarray}
where, also, $|v_{\rm o}| \ll |u_{\rm o}|$ together with $|K_{\rm o}| = 1$ have been used. It may be noted that the latter approximation cannot be used in the limit  $|\delta k|\tau < 1$, since all of the first order terms cancel, and the expression for $|K_{\rm o}|^2 -1$ is important in order to get the correct second order terms. With
\begin{equation}
	\frac{\alpha}{|\alpha|^2} = \frac{2\delta k \chi^{*}}{\tau|\delta k|^2 },
	\label{eq24}
\end{equation}	
the various contributions to the circular polarisation can be directly estimated from Equation (\ref{eq23}). One may note that $|\alpha|^{-1} \sim |\chi|/|\delta k| \tau$, which is useful when estimating the effects of coupling relative the WKB-term. 

Consider first the case $|\rho| \ll 1$ (nearly linear characteristic waves), for which $u_{\rm hom}(\tau) = u_{\rm o}\cosh(\delta k_{\rm r}\tau) + \sinh (\delta k_{\rm r}\tau),  K_{\rm r} = -1$ and $\delta k = -\xi_{\rm U} + i\hat{\xi}_{\rm U}$ \citep{bjo19}. Since $u_{\rm o} \approx 1$, it is seen from Equation (\ref{eq24}) that the WKB-term $\chi_{\rm i} u_{\rm hom}$ in Equation (\ref{eq23}) is a factor $\sim |\delta k| \tau$ ($> 1$) larger than the coupling terms; hence,
\begin{equation}
	V = I_{\rm o}\exp(-\tau)[v_{\rm hom} - 2\chi_{\rm i}\{u_{\rm o}\cosh(\delta k_{\rm r}\tau) + \sinh (\delta k_{\rm r}\tau)\}].
	\label{eq25}
\end{equation}
Furthermore, $v_{\rm hom} = K_{\rm o,i}\{u_{\rm o}\cosh(\delta k_{\rm r}\tau) + \sinh (\delta k_{\rm r}\tau)\}$ and $2\chi_{\rm i} = K_{\rm o,i} - 
K_{\rm i}$, which give the simple expression
\begin{equation}
	V = I_{\rm o}\exp(-\tau)K_{\rm i}\{u_{\rm o}\cosh(\delta k_{\rm r}\tau) + \sinh (\delta k_{\rm r}\tau)\},
	\label{eq26}
\end{equation}
or  $V = (K_{\rm i}/K_{\rm o,i})V_{\rm hom}$. $K_{\rm i}$ is the polarisation of the characteristic waves appropriate for the surface and, hence, is independent of $\tau$ for a given line of sight. One may note that in this limit, none of the terms $\propto |\alpha|^{-1}$ in Equation (\ref{eq23}) contributes to the circular polarisation. Hence, this result corresponds to the WKB-approximation discussed in Section\,\ref{sect2}. Since $K_{\rm i} \approx \hat{\xi} _{\rm V}/\hat{\xi} _{\rm U} \propto \nu/B$, where $B$ is the strength of the magnetic field, the inhomogeneities change the value of the circular polarisation by a factor $B_{\rm o}/B$. The main conclusions for nearly linear characteristic waves are then: (1) The inhomogeneities can substantially affect the circular polarisation at frequencies for which $|\delta k|\tau \gsim 1$. (2) The amplitude of the change is independent of frequency, and hence, the frequency distribution of the circular polarisation remains the same as for a homogeneous source. Both of these features can be seen explicitly in the numerical solutions to the transport equation shown in \cite{bjo90}.

For $|\rho| \gg 1$ (nearly circular characteristic waves), $u_{\rm hom}(\tau) = u_{\rm o}\cos(\delta k_{\rm i}\tau) + q_{\rm o}\sin (\delta k_{\rm i}\tau)$, $K_{\rm i} = 1$ and  $\delta k  =  -(\xi_{\rm V} + \xi_{\rm U}\hat{\xi}_{\rm U}/\hat{\xi}_{\rm V})+ i\hat{\xi}_{\rm V}$. As already discussed, integration along the line of sight is expected to give $u_{\rm hom} \approx 0$ so that 
\begin{eqnarray}
	V &=& I_{\rm o}\exp(-\tau)\times\nonumber\\ 
	&&\left[v_{\rm hom} + \frac{2u_{\rm o}(\delta k_{\rm r}\chi_{\rm i} - \delta k_{\rm i}\chi_{\rm r})}{|\delta k|}\frac{\sinh (\delta k_{\rm r}\tau)}
	{|\delta k| \tau} + \frac{2q_{\rm o}(\delta k_{\rm r}\chi_{\rm r} + \delta k_{\rm i}\chi_{\rm i})}{|\delta k|}\frac{\cosh (\delta k_{\rm r}\tau)}
	{|\delta k| \tau}\right].
	\label{eq27}
\end{eqnarray}
Since $|\hat{\xi}_{\rm V}| \gg 1$ and $|\delta k_{\rm r}| \ll 1$, the magnitude of the two inhomogeneous terms are roughly $|u_{\rm o}\chi_{\rm r}\delta k_{\rm r}/\delta k_{\rm i}|$ and $|q_{\rm o}\chi_{\rm i}/(|\delta k_{\rm i}|\tau)|$. As compared to $|v_{\rm hom}| \sim u_{\rm o}|\hat{\xi}_{\rm U}/\hat{\xi}_{\rm V}|$, the first term is smaller by a factor $|\hat{\xi}_{\rm v}/\chi_{\rm r}| \gg 1$, while the second one could be of the same order of magnitude. However, for synchrotron radiation, $q_{\rm o} = 0$. Hence, for nearly circular characteristic waves, the emerging polarisation is given to a good approximation by the homogeneous value, and the inhomogeneities will affect the result only marginally.

\subsection{Magnetic field with a varying azimuthal angle}\label{sect3b}
When the azimuthal angle varies along the ray path, the expressions for $K^1$ and $K^2$ lead to a more complex relation between $\Psi^1$ and $\Psi^2$. This is caused by the term $\propto \sin(2\varphi)$ in Equation (\ref{eq4}). In general, under such conditions, no simple solutions can be found for the transport equation. However, as discussed in Section\,\ref{sect3a}, the low value of the circular polarisation observed in compact radio sources suggests that the relative variations of $K^1$ and $K^2$ along the ray path is rather small. Defining
\begin{equation}
	K^{1,2} \equiv K_{\rm o}(\mp1 + 2\chi^{1,2})
	\label{eq28}
\end{equation}
and assuming $|\chi^{1,2}| \ll1$, one finds to lowest order that
\begin{eqnarray}
	\chi^{1,2} &=& \chi_{\rm +} \pm \chi_{\rm -}\nonumber\\
	&=& \frac{\sin(2\varphi)}{2\sqrt{1-\rho_{\rm o}^2}} \pm \left\{\frac{\Delta \rho}{2(1-\rho_{\rm o}^2)} - \frac{1- \cos(2\varphi)}
	{2(1+\rho_{\rm o})}\right\}.
	\label{eq29}
\end{eqnarray}
Here, $\rho = \rho_{\rm o} + \Delta \rho$ and, again, $K_{\rm o} \equiv K_{\rm o}^2 = -\sqrt{(1-\rho_{\rm o})/(1+\rho_{\rm o})}$. When, $|\chi_{\rm +}| \ll |\chi_{\rm -}|$, $\Psi^1 = \Psi^2$, while $|\chi_{\rm +}| \gg |\chi_{\rm -}|$ leads to $\Psi^1 = -\Psi^2$. Since the former situation is treated in Section\,\ref{sect3a}, the focus in this section is on the latter case.

From Equation (\ref{eq28}) one deduces
\begin{equation}
	\exp \int_0^{s}\frac{{\rm d}K^{1,2}}{{\rm d}\hat{s}}\frac{{\rm d}\hat{s}}{(K^{2,1} - K^{1,2})} = 1 \pm \chi_{\rm +}.
	\label{eq32}
\end{equation}
Equation (\ref{eq17}) then leads to
\begin{eqnarray}
	E_{\rm x} =E_{\rm x}^1 + E_{\rm x}^2 & = &  \left\{\tilde{E}^1 + \tilde{E}^2 + \chi_{\rm +}(\tilde{E}^1 - \tilde{E}^2)\right\}
	\exp(-\tau/2)\nonumber\\
	E_{\rm y} =K^1 E_{\rm x}^1 + K^2 E_{\rm x}^2 & = &  -K_{\rm o}\left\{\tilde{E}^1 - \tilde{E}^2 -\chi_{\rm +}(\tilde{E}^1 + \tilde{E}^2)\right\}
	\exp(-\tau/2).
	\label{eq33}
\end{eqnarray}
Furthermore, the coupling constants (see Equation \ref{eq14}) are given by $\Psi^{1,2} = \mp {\rm d} \chi_{\rm +}/{\rm ds}$. In order to emphasise the similarities to Section\,\ref{sect3a}, the same notation will be used; hence, $\chi \equiv \chi_{\rm +}$ and
$\alpha = \Delta k /(2{\rm d}\chi/{\rm d}s)$. Furthermore, with $\alpha = constant$, it can be expressed as $\alpha = \delta k \tau/2\chi$. The equations for $X$ and $Y$ are then
\begin{eqnarray}
	\frac{{\rm d}X}{{\rm d}\chi} & = &  (\alpha - 1)Y
	\nonumber\\
	\frac{{\rm d}Y}{{\rm d}\chi} & = & (\alpha + 1)X .
	\label{eq35}
\end{eqnarray}
These equations resemble those in Section\,\ref{sect3a} and can be solved in a similar manner. This is done in Appendix B. The solution is in this case
\begin{eqnarray}
	E_{\rm x} = \sqrt{\frac{I_{\rm o}}{2(1+q_{\rm o})}}&\times&\left[(1+q_{\rm o})\cosh(\beta\chi)+(1 - \alpha)\sigma_{\rm o}
	\frac{\sinh(\beta\chi)}{\beta}\right.
	\nonumber\\ 
	&+&\left.\chi\left\{-\sigma_{\rm o}\cosh(\beta\chi) + (1+\alpha)(1+q_{\rm o})\frac{\sinh(\beta\chi)}{\beta}\right\}\right]
	\exp(-\tau/2)\nonumber\\
	E_{\rm y} =\sqrt{\frac{I_{\rm o}}{2(1+q_{\rm o})}}&\times&K_{\rm o}\left[\sigma_{\rm o}\cosh(\beta\chi) - (1 + \alpha)(1+q_{\rm o})
	\frac{\sinh(\beta\chi)}{\beta}\right.\nonumber\\ 
	&+&\left.\chi\left\{(1+q_{\rm o})\cosh(\beta\chi) + (1-\alpha)\sigma_{\rm o}\frac{\sinh(\beta\chi)}{\beta}\right\}\right]
	\exp(-\tau/2),\nonumber\\
	\label{eq36}
\end{eqnarray}
where, now, $\beta = \sqrt{\alpha^2-1}$. 

Expansion of Equation (\ref{eq36}) to lowest order in $|\beta \chi|$ gives
\begin{eqnarray}
	E_{\rm x} &=& \sqrt{\frac{I_{\rm o}}{2(1+q_{\rm o})}}\left[1+q_{\rm o} + \frac{\delta k \tau}{2}\{ - \sigma_{\rm o} + (1+q_{\rm o}) \chi\}\right]
	\exp(-\tau/2)\nonumber\\
	E_{\rm y} &=&\sqrt{\frac{I_{\rm o}}{2(1+q_{\rm o})}}K_{\rm o}\left[\sigma_{\rm o} - \frac{\delta k \tau}{2}\{1+q_{\rm o} +\sigma_{\rm o}\chi
	\}\right]\exp(-\tau/2),
	\label{eq37}
\end{eqnarray}
which leads to the simple expression
\begin{eqnarray}
	U+iV \equiv 2E_{\rm x}E_{\rm y}^{*} = I_{\rm o}\exp(-\tau)[u_{\rm hom} + iv_{\rm hom} + (u_{\rm o} + iv_{\rm o})i \tau
	(\delta k \chi)_{\rm i}].
	\label{eq38}
\end{eqnarray}
Furthermore, since $\delta k \chi = \delta k \sin(2\varphi)/(2\sqrt{1-\rho_{\rm o}^2})$ and $\varphi$ is real, one finds from Equation (\ref{eq38})
\begin{equation}
	V = I_{\rm o}\exp(-\tau)[v_{\rm hom} + u_{\rm o}\hat{\xi}_{\rm U}\tau \sin(2\varphi)/2],
	\label{eq39}
\end{equation}
which is valid for both $|\rho| \ll 1$ and $|\rho| \gg 1$. Similarly to the situation in Section\,\ref{sect3a}, the first order terms vanish, and the impact of the inhomogeneities is given by the second order term $\tau\delta k \chi$. However, in contrast to the $\Psi^1 = \Psi^2$ case, the inhomogeneous term here cannot be neglected. 

Equation (\ref{eq39}) is valid for $|\beta \chi| \approx |\delta k| \tau < 1$. For $|\rho| \ll 1$, this implies that in the transition region (i.e., $|\hat{\xi}_{\rm U}| \tau \sim 1$), the inhomogeneous term approaches the value $u_{\rm o} \varphi $. This should be compared to the corresponding value of the homogeneous term, which is $ \sim u_{\rm o}\hat{\xi}_{\rm V}/\hat{\xi}_{\rm U}$. Since $|\hat{\xi}_{\rm V}/\hat{\xi}_{\rm U}| \ll 1$ in this limit, the inhomogeneities could dominate the circular polarisation. Likewise, for $|\rho| \gg 1$, the value in the transition region (i.e., $|\hat{\xi}_{\rm V}| \tau \sim 1$) is $\sim u_{\rm o} (\hat{\xi}_{\rm U}/\hat{\xi}_{\rm V})(\sin(2\varphi)/2)$, which should be compared to the corresponding expression for the homogeneous part,  $\sim u_{\rm o} \hat{\xi}_{\rm U}/\hat{\xi}_{\rm V}$. Since $|\varphi| \sim 1$ is allowed in this limit, inhomogeneities may significantly affect the circular polarisation in the transition region also in this case, although the change in $\varphi$ must be much larger than for $|\rho| \ll 1$.

In analogy with Section\,\ref{sect3a}, it is useful to expand Equation (\ref{eq36}) to first order in $|\alpha|^{-1}$, which yields (see Appendix B)
\begin{eqnarray}
	U + iV &=& I_{\rm o} \exp(-\tau)[u_{\rm hom} + iv_{\rm hom} + 2i(u_{\rm o} + iv_{\rm o})\left\{\chi \cosh^{*}(\delta k \tau/2)
	\sinh(\delta k \tau/2)\right\}_{\rm i}\nonumber\\
	&+&2i\sigma_{\rm o} K_{\rm o}^{*}\left\{\frac{\alpha}{|\alpha|^2}|\sinh(\delta k \tau/2)|^2 + \chi \cosh(\delta k \tau/2)
	\sinh^{*}(\delta k \tau/2)\right\}_{\rm i}\nonumber\\
	&-&2iK_{\rm o}^{*}\left\{\chi_{\rm i} \cosh(\delta k_{\rm r} \tau) + \frac{\{\alpha \cosh(\delta k \tau/2)\sinh^{*}
	(\delta k \tau/2)\}_{\rm i}}{|\alpha|^2}\right\}\nonumber\\
	&+&\left.2q_{\rm o}K_{\rm o}^{*}\left\{\chi_{\rm r} \cos(\delta k_{\rm i} \tau) - \frac{\{\alpha \cosh(\delta k \tau/2)\sinh^{*}
	(\delta k \tau/2)\}_{\rm r}}{|\alpha|^2}\right\}\right].
	\label{eq40}
\end{eqnarray}
Equation (\ref{eq40}) is written so as to highlight the cancellations that occur for $|\delta k|\tau < 1$; namely, the last three inhomogeneous terms vanish to second order (i.e., $|\delta k \tau \chi|$), so that the result in Equation (\ref{eq39}) corresponds to the first term only. On the other hand, all of the terms contribute in the limit $|\delta k|\tau > 1$,
\begin{eqnarray}
	U + iV &=& I_{\rm o} \exp(-\tau)[u_{\rm hom} + iv_{\rm hom} + i(u_{\rm o} + iv_{\rm o})\chi_{\rm i}
	\sinh(\delta k_{\rm r} \tau)\nonumber\\
	&+&i\sigma_{\rm o} K_{\rm o}^{*}\left\{\frac{\alpha_{\rm i}}{|\alpha|^2}\cosh(\delta k_{\rm r} \tau) + \chi_{\rm i} \sinh(\delta k_{\rm r} \tau)
	\right\}\nonumber\\
	&-&iK_{\rm o}^{*}\left\{2\chi_{\rm i} \cosh(\delta k_{\rm r} \tau) + \frac{\alpha_{\rm i}}{|\alpha|^2}\sinh(\delta k_{\rm r}\tau)\right\}
	\nonumber\\
	&-&\left.q_{\rm o}K_{\rm o}^{*}\frac{\alpha_{\rm r}}{|\alpha|^2}\sinh(\delta k_{\rm r} \tau)\right],      
	\label{eq41}
\end{eqnarray}
where, again, terms $\propto \cos(\delta k_{\rm i} \tau)$ and $\propto \sin(\delta k_{\rm i} \tau)$ have been neglected.

Consider first the case $|\rho| \ll 1$ ($K_{\rm o} = -1$ and $\chi$ real). The circular polarisation obtained from Equation (\ref{eq41}) is then for a synchrotron source ($q_{\rm o}= 0$),
\begin{equation}
	V = I_{\rm o} \exp(-\tau)\left[v_{\rm hom} + \frac{\alpha_{\rm i}}{|\alpha|^2}\{u_{\rm o}\cosh(\delta k_{\rm r} \tau) + 
	\sinh(\delta k_{\rm r} \tau)\}\right].
	\label{eq42}
\end{equation}
With the use of Equation (\ref{eq24}), this can be written
\begin{equation}
	V = I_{\rm o} \exp(-\tau)\left[v_{\rm hom} + \frac{2\varphi \hat{\xi}_{\rm U}}{\tau (\xi_{\rm U}^2 +\hat{\xi}_{\rm U}^2)}
	\{u_{\rm o}\cosh(\xi_{\rm U}\tau) - \sinh(\xi_{\rm U}\tau)\}\right],
	\label{eq43}
\end{equation}
where $\delta k = -\xi_{\rm U} + i \hat{\xi}_{\rm U}$ has been used. 

Likewise, for  $|\rho| \gg 1$ ($K_{\rm o} = i, |\hat{\xi}_{\rm V}| \gg 1$ and $|\xi_{\rm V}| \ll 1$),
 \begin{equation}
	V = I_{\rm o} \exp(-\tau)\left[v_{\rm hom} - u_{\rm o}\frac{\alpha_{\rm i}}{|\alpha|^2}\cosh(\delta k_{\rm r} \tau) \right].
	\label{eq44}
\end{equation}
Again, Equation (\ref{eq24}) can be used to find
\begin{equation}
	V = I_{\rm o} \exp(-\tau)\left[v_{\rm hom} - u_{\rm o}\frac{\sin(2\varphi) \hat{\xi}_{\rm U}}{\tau \hat{\xi}_{\rm V}^2}
	\cosh(\xi_{\rm U}\hat{\xi}_{\rm U}\tau/\hat{\xi}_{\rm V})\right],
	\label{eq45}
\end{equation}
where $\delta k = -\xi_{\rm U}\hat{\xi}_{\rm U}/\hat{\xi}_{\rm V} + i\hat{\xi}_{\rm V}$ and $\delta k_{\rm r}/\delta k_{\rm i} \approx |\hat{\xi}_{\rm V}|^{-2} \ll1 $ have been used.

It is seen that in the transition region (i.e., $|\delta k| \tau \sim 1$), the inhomogeneities induce a circular polarisation $\sim u_{\rm o}|\chi|$ (Equation \ref{eq38}). As discussed above, this may correspond to a significant fraction of the homogeneous value. When this is the case, the transition between the regimes $|\delta k| \tau < 1$ and $|\delta k| \tau > 1$ is smoother than the more abrupt one expected for $\Psi^1 = \Psi^2$ (see discussion in Section\,\ref{sect3a}). Even so, since the frequency dependence of $\hat{\xi}_{\rm U}$ is rather weak, the increase in circular polarisation toward lower frequencies is still rather steep, $V\,\propsim\,\tau\,\propsim\,\nu^{-3}$ (see Equation \ref{eq39}). It should also be noted that the frequency range over which the circular polarisation is enhanced by inhomogeneities is rather narrow, since it declines as $V \propto \tau^{-1}$ for $|\delta k| \tau > 1$. In addition, for $|\rho| \ll 1$, this frequency range is further narrowed down by the fact that the circular polarisation changes sign not too far from the transition region (cf. Equation \ref{eq43}), while for $|\rho| \gg 1$, it is somewhat broadened by the frequency dependence of $\hat{\xi}_{\rm V}$ ($\tau \hat{\xi}_{\rm V}^2\,\propsim\,\nu^{-1}$; see Equation \ref{eq45}). The detailed spectral properties of the circular polarisation can be seen in \cite{hod82}, who solved the transport equation numerically for $\rho = 0$.

\section{Comparison with a piecewise constant approximation}\label{sect4}
The validity of the approximation used in Section\,\ref{sect3} constrains the allowed variation of $\alpha$ (see Appendix A). One may note that this $\alpha = constant$ description has much wider applicability than the WKB-approximation discussed in Section\,\ref{sect2}, since the latter corresponds to $\alpha \rightarrow \infty$ (i.e., no coupling between the characteristic waves). The smoothly varying inhomogeneities, assumed in the $\alpha = constant$ description, can be contrasted by the piecewise constant approximation in which inhomogeneities are modelled as instantaneous changes of the plasma properties interspersed by homogeneous regions \citep[e.g.,][]{r/b02,m/m18}. The instantaneous changes do not affect the polarisation, since they corresponds to $\alpha \rightarrow 0$. These two approximations are each others opposites, i.e., mutually exclusive. It is, therefore, useful to compare their results for a given situation. This will give an estimate of the sensitivity of the emerging polarisation to the approximation used to calculate it.

Consider a region of length $s$, where the azimuthal angle of the magnetic field changes by $\varphi$. Assume further that ${\rm d}\phi / {\rm d}s$ is constant, which implies $\varphi = s {\rm d}\phi / {\rm d}s$. Let $s$ be small enough so that $|\delta k \tau| = |\delta k| \kappa s  \ll 1$ and $|\varphi| \ll 1$. Divide the region in two, with $s = s_{\rm 1} + s_{\rm 2}$ and $\varphi =\varphi_{\rm 1} + \varphi_{\rm 2}$.  For a light ray emitted at $s=0$, one finds from Equation (\ref{eq39}) that at the end of region $1$,
\begin{equation}
	v_{\rm 1}^{\alpha} = v_{\rm hom} + u_{\rm o}\hat{\xi}_{\rm U}\kappa \frac{{\rm d}\phi}{{\rm d} s}s_{\rm 1}^2,
	\label{eq46}
\end{equation}
where, for convenience, $v_{\rm 1}^{\alpha} \equiv V/(I_{\rm o} \exp (-\tau))$ has been introduced. For synchrotron radiation, $q_{\rm o} = 0$ can be chosen at the beginning of region $1$. However, at the beginning of region $2$, the rotation of the magnetic field implies $q_{\rm o} = -2\varphi_{\rm 1}u_{\rm o}$, while, to first order in $\varphi_{\rm 1}, u_{\rm o}$ remains the same. With the use of Equation (\ref{eq39}) a second time, the circular polarisation at the end of region $2$ can be written
\begin{eqnarray}
	v_{\rm 2}^{\alpha} &=& v_{\rm 1}^{\alpha} + u_{\rm o}\hat{\xi}_{\rm U}\kappa \frac{{\rm d}\phi}{{\rm d} s}s_{\rm 2}^2 - 
	q_{\rm o}\hat{\xi}_{\rm U}\kappa s_{\rm 2}\nonumber\\
	&=& v_{\rm hom}  + u_{\rm o}\hat{\xi}_{\rm U}\kappa \frac{{\rm d}\phi}{{\rm d} s}(s_{\rm 1} + s_{\rm 2})^2,
	\label{eq47}
\end{eqnarray}
where the term $\propto q_{\rm o}$ comes from the homogeneous solution (cf. Equation {\ref{eq21b}}). Hence, the expression for $v_{\rm hom}$ in Equation (\ref{eq47}) is that for a homogeneous source with $q_{\rm o} = 0$, just as in Equation (\ref{eq46}). It is seen that the circular polarisation is additive, as it should be, since a given region can be divided up in any number of subregions without affect the resulting value of $V$.

If the same region is approximated by a piecewise constant medium, there will be two abrupt changes in the azimuthal angle of the magnetic field, one at the beginning of region $1$ and a second one at the beginning of region $2$, given by $\varphi_{\rm 1}$ and $\varphi_{\rm 2}$, respectively. The equations corresponding to Equations (\ref{eq46}) and (\ref{eq47}) are
\begin{equation}
	v_{\rm 1}^{pw} = v_{\rm hom} + 2\varphi_{\rm 1}u_{\rm o}\hat{\xi}_{\rm U}\kappa s_{\rm 1},
	\label{eq48}
\end{equation}
and 
\begin{eqnarray}
	v_{\rm 2}^{pw} &=& v_{\rm 1}^{pw} + 2(\varphi_{\rm 1} + \varphi_{\rm 2})u_{\rm o}\hat{\xi}_{\rm U}\kappa s_{\rm 2}\nonumber\\
	&=& v_{\rm hom}  + 2u_{\rm o}\hat{\xi}_{\rm U}\kappa \frac{{\rm d}\phi}{{\rm d} s}((s_{\rm 1} + s_{\rm 2})^2 - s_{\rm 1}s_{\rm 2}),
	\label{eq49}
\end{eqnarray}
It is seen that the transport induced circular polarisation calculated with the use of the two different approximations differs by a factor
\begin{equation}
	\frac{v_{\rm 2}^{pw} - v_{\rm hom}}{v_{\rm 2}^{\alpha} - v_{\rm hom}} = 2\left(1-\frac{s_{\rm 1}s_{\rm 2}}{(s_{\rm 1} + s_{\rm 2})^2}
	\right).
	\label{eq50}
\end{equation}
Hence, depending on how the division of the region is done, the piecewise constant approximation can give circular polarisation up to a factor of two larger than the $\alpha$-approximation. Likewise, in situations where the main changes of the magnetic field are so abrupt that the $\alpha$-approximation is not applicable, its use would give an artificially low circular polarisation.

Another aspect of the two approximations is how they account for variations in the sign of ${\rm d}\phi /{\rm d}s$ along a ray path. In order to illustrate this, let the sign of ${\rm d}\phi /{\rm d}s$ change between region $1$ and $2$ (i.e., $s_{\rm 2} 
\rightarrow -s_{\rm 2}$ in the equations above). This lowers the value of $v_{\rm 2}^{\alpha} $. The reason is that the value of $\varphi$ in the $\alpha$-approximation is the integrated change of the azimuthal angle of the magnetic field along a ray path; e.g., $s_{\rm 2} = s_{\rm 1}$ gives $v_{\rm 2}^{\alpha} -v_{\rm hom} = 0$. Hence, the detailed properties of the medium along a ray path can vary substantially without affecting the emerging polarisation. This is not so for the piecewise constant approximation; for example, with $s_{\rm 2} = s_{\rm 1}$ one finds from Equation (\ref{eq49})
\begin{equation}
	v_{\rm 2}^{pw} = v_{\rm hom}  + 2u_{\rm o}\hat{\xi}_{\rm U}\kappa \frac{{\rm d}\phi}{{\rm d} s}s_{\rm 1}^2.
	\label{eq51}
\end{equation}
This shows explicitly that in this case, the circular polarisation is also sensitive to the detailed properties of the medium.

The polarisation of the emerging radiation is obtained by adding up all the light rays along the line of sight. Hence, the circular polarisation results from a combination of an integration over the initial conditions of the light rays and their propagation through the medium. Consider, for example, a turbulent medium in which the sign of ${\rm d}\phi /{\rm d}s$ changes repeatedly along a sight line. Although the circular polarisation for a given light ray is likely to be quite different depending on whether the $\alpha$-approximation or the piecewise constant approximation is used, both approximations are $\propto {\rm d}\phi /{\rm d}s$ so the statistical properties of the medium will affect them in a similar manner. The relative importance of the two effects depends on the detailed properties of the medium. However, one may expect the larger fluctuations between different light rays in the piecewise constant approximation to give rise to a higher circular polarisation than that resulting from the $\alpha$-approximation. 

\section{Discussion}\label{sect5}
Before addressing the observed polarisation of compact radio sources, it is useful to discuss a few general properties of the transport equation for polarised light. Normally, this equation is expressed using the Stokes parameters. Alternatively, it can be written in terms of the electric field. In this latter formulation, the Stokes parameters are then calculated from the solution to the transport equation.

As argued in \cite{bjo19}, the interaction between the electromagnetic wave and the plasma is more transparently described in terms of the electric field directly rather than via the Stokes parameters. This is particularly true when the concept of characteristic waves is introduced. In a homogeneous source, these waves propagate independently and allow both a straightforward solution to the transport equation as well as a simple formulation of the full result in terms of the polarisation properties of the characteristic waves.

In the inhomogeneous case, there exists a WKB-approximation, which, physically, corresponds to negligible coupling between the characteristic waves. Instead, the polarisation of the propagating characteristic waves change in tune with the local properties of the plasma along a given ray path. However, this solution has limited applicability. The general description of the effects of inhomogeneities is instead formulated in terms of the coupling between the characteristic waves.

The standard derivation of the equations accounting for this coupling is rather tedious. Furthermore, it is normally written in a form that is not so physically transparent. A shorter and more straightforward derivation is presented in Section\,\ref{sect2}. In addition, the equations can be written in a way so as to highlight the main physical effects. Most importantly, as shown in Section\,\ref{sect3}, these equations have a constant coupling solution ($\alpha = constant$) with an applicability much wider than the WKB-approximation; for example, the latter is recovered in the limit of no coupling (i.e., $\alpha \rightarrow \infty$). When this $\alpha$-approximation is valid, the calculation of the emerging polarisation is much simplified; instead of solving coupled differential equations, one needs only to integrate over the conditions along a given line of sight. 

This integration consists of two parts. The constant transport coefficients in the homogeneous case are substituted by their average values appropriate for a given light path. After this, the total polarisation is obtained by integrating over the varying initial conditions along the line of sight. Furthermore, the solution for a given light ray can be represented as the sum of two terms. The first corresponds to the solution for the homogeneous case but with the constant phase difference between the characteristic waves substituted by their average value, while the second one accounts for the varying polarisation properties of the characteristic waves. Moreover, it is shown that under a rather wide range of circumstances, the first, "homogeneous" term dominates the resulting  polarisation. 

Another approximation sometimes used is based on the assumption of a piecewise constant medium, in which a given variation of the plasma properties is modelled as an instantaneous change followed by a homogeneous region \citep{r/b02,m/m18}. This is in contrast to the $\alpha$-approximation, which relies on smooth variations. The validity of these two approximations do not overlap, and hence, they apply to very different situations. As shown in Section\,\ref{sect4}, a given change of plasma properties can result in a variation of in the value of the circular polarisation differing by up to a factor two, depending on which of the approximations is used. Furthermore, while the $\alpha$-solution is expressed in terms of integrated properties along the ray path only, the result from the piecewise constant approximation is more sensitive to the local properties of the plasma.

\subsection{Polarisation properties of compact radio sources}\label{sect5a}
The low value of the circular polarisation observed in compact radio sources makes it likely that the characteristic waves are either nearly linearly polarised or nearly circularly polarised. The main aim of the present paper is to use the observed polarisation to distinguish between the two. In \cite{bjo19}, the polarisation properties of a homogeneous synchrotron source were used to argue that the properties of compact radio sources are such that the characteristic waves are nearly circularly polarised. In addition, qualitative arguments were given as to why this may also apply to inhomogeneous sources. Here, a quantitative estimate is done of the effects that inhomogeneities may have on the polarisation emerging from a synchrotron source.

In \cite{bjo19}, it was shown that the two types of characteristic waves result in very different frequency dependences of the circular polarisation. This difference is due to the relative values of the circular and linear birefringence ($\hat{\xi}_{\rm V}$ and $\hat{\xi}_{\rm U}$, respectively) for the two types. For nearly circular characteristic waves ($|\hat{\xi}_{\rm V}/\hat{\xi}_{\rm U}| \gg 1$ and $|\hat{\xi}_{\rm V}| \gg 1$), most of the circular polarisation is emitted over a rather wide range of optically thin frequencies ($|\hat{\xi}_{\rm V}|^{-1}\lsim\,\tau\,\lsim 1 $). In contrast, for nearly linear characteristic waves ($|\hat{\xi}_{\rm U}/\hat{\xi}_{\rm V}| \gg 1$ and $|\hat{\xi}_{\rm U}|\,\lsim 1$), the circular polarisation is emitted over a rather narrow range of mainly optically thick frequencies ($\tau\,\gsim\,|\hat{\xi}_{\rm U}|^{-1}$).

There are two instances that give rise to nearly linear characteristic waves. Firstly, when the lower cut-off in the energy distribution of the relativistic electrons corresponds to synchrotron frequencies close to self-absorption and, secondly, the presence of a substantial amount of electron-positron pairs. Likewise, nearly circular characteristic waves imply a rather small value for the low energy cut-off in the energy distribution. For example, consider the spectral region around the synchrotron self-absorption peak, where most of the circularly polarised flux is emitted. Here, $\hat{\xi}_{\rm V}/\hat{\xi}_{\rm U} \sim 10^4/(\gamma_{\rm min}^3(1+2\hat{n}))$, where $\gamma_{\rm min}$ is the lower cut-off in the energy distribution of the relativistic electrons and $\hat{n}$ is the number of pairs per proton \citep[Appendix C in][]{bjo19}. 

It is shown in Section\,\ref{sect3} that inhomogeneities affect the emerging circular polarisation substantially more for nearly linear as compared to nearly circular characteristic waves. However, the most important point for this paper is that the main influence of the inhomogeneities is restricted to the amplitude of the circular polarisation; its frequency dependence is only marginally affected. Hence, the conclusion in \cite{bjo19}, that the observed properties of the polarisation in compact radio sources are most directly understood as the result of nearly circular characteristic waves, also remains valid in the presence of inhomogeneities. The observations then imply $\hat{\xi}_{\rm V}/\hat{\xi}_{\rm U} \sim 10^2$ or $\gamma_{\rm min}^3(1+2\hat{n}) \sim 10^2$. The degeneracy between $\gamma_{\rm min}$ and $\hat{n}$ may be broken by observations at frequencies for which $\tau\,\lsim\,|\hat{\xi}_{\rm V}|^{-1}$. As argued in Section\,\ref{sect3}, in this frequency range, the observed circular polarisation is likely dominated by the emission process itself, which is independent of $\gamma_{\rm min}$ but inversely proportional to $\hat{n}$.

In the POLAMI survey \citep{thu17}, a sample of compact radio sources were observed multiple times at 1.3\,mm and 3\,mm. Except for some periods of increasing flux, the spectral index indicated that the emission was optically thin. Furthermore, the degrees of circular polarisation at the two wavelengths were rather similar. This is consistent with nearly circular characteristic waves but hard to reconcile with nearly linear characteristic waves, since the latter are expected to show a steeply rising degree of circular polarisation toward longer, optically thin wavelengths. Although it is shown in Section\,\ref{sect3b} that inhomogeneities may smooth this very steep rise for a homogeneous source, it is still hard to make the expected rise ($V\,\propsim\,\nu^{-3}$) compatible with observations. 

The degree of circular polarisation at lower frequencies, where the flat spectra indicate optically thick emission, is smaller than that observed in the POLAMI survey. Both types of characteristic waves show a sign-change of the circular polarisation at optically thick frequencies. As discussed in \cite{bjo19}, the relative magnitude of this contribution is substantially larger for nearly circular as compared to nearly linear characteristic waves. Since the flat spectra are likely due to an inhomogeneous jet \citep{b/k79}, the emission at a given frequency is that obtained by integrating over a range of optical depths. Hence, the degree of circular polarisation can be lowered by contributions from optically thick regions with different signs of the circularly polarised flux. For nearly linear characteristic waves, this effect is quite small, while for nearly circular characteristic waves, it can be substantial. The observed clear decrease of circular polarisation when going from optically thin to thick frequencies favours the presence of nearly circular characteristic waves. This will be discussed in more detail in a forthcoming paper. 

When transport effects are important, the choice between nearly linear and nearly circular characteristic waves relies not only on the observed properties of the circular polarisation but also on those of the flux and linear polarisation are important as well. This is so because the polarisation of the emerging radiation is determined by the low energy electrons, which may be different from those giving rise to the bulk of the flux. Since the value of $\hat{\xi}_{\rm U}$ is independent of $\hat{n}$ and varies slowly with $\gamma_{\rm min}$, it is the value of $\hat{\xi}_{\rm V}$ that distinguishes between the two types of characteristic waves. In principal then, determination of the amount of Faraday rotation alone would settle the issue. However, for an inhomogeneous source, this is not straightforward. 

As discussed in \cite{bjo19}, the observed properties of linear polarisation in flat spectrum radio sources can be understood as the result of large Faraday depths. The longer timescale of variability for the linear polarisation, as compared to the circular polarisation, would be due to an emission site further out in the optically thin part of the jet ($\tau\,\lsim |\,\hat{\xi}_{\rm V}|^{-1}$). At the same time, this would lower the degree of linear polarisation. Quantitatively, both of these effects are consistent with observations for $|\hat{\xi}_{\rm V}|\sim 10^2$.  

Large Faraday depths imply different frequency distributions for the circular and linear polarisation. Since the circularly polarised flux comes mainly from the region close to the spectral peak, while the linearly polarised flux comes from the optically thin part of the spectrum where $\tau\,\lsim |\,\hat{\xi}_{\rm V}|^{-1}$, a broad minimum of the linear polarisation is expected in the frequency range where the circular polarisation peaks. In flat spectrum radio sources, the spectral peak usually occurs at $\sim$100\,GHz. At such large frequencies, the spectral resolution is not yet sufficient to establish the presence of such an anti-correlation between linear and circular polarisation. The situation is different for Gigahertz-Peaked-Spectrum sources where the spectrum peaks at $\sim$\,few GHz. For such sources, high-quality, multifrequency observations are now possible. A good example is PKS B2126-158 \citep{osu13}, which shows a clear anti-correlation between circular and linear polarisation as expected for nearly circular characteristic waves.

\subsection{Implications for the acceleration process}\label{sect5b}
Numerical calculations based on first principles are now possible for the acceleration of particles. Although limited in scope, they are likely to give realistic insights to the injection of particles and their low energy distribution. Hence, the observed properties of the circular polarisation should provide useful constraints for the results from such PIC-simulations. In order to illustrate this, consider the distribution of electron energies calculated in \cite{pet19} from magnetic reconnection. In general, a rising thermal tail is followed by a decreasing, roughly power-law distribution at higher energies, where the peak normally falls in a region around a Lorentz factor $\gamma_{\rm p} \sim10$. \cite{j/h79} have shown that the transport coefficients for a relativistic Maxwellian are dominated by the particles around the peak energy, i.e., that the low energy tail contributes negligibly \citep[see also][]{bjo90}. Hence, the transport coefficients obtained from the energy distributions calculated in \cite{pet19} should be approximately those of a power-law distribution with a low-energy cut-off at $\gamma_{\rm p}$. 

It has been argued that observations constrain the number of electron-positron pairs per proton to be around $10$ \citep{ghi10,maj16}. Together with $\gamma_{\rm p} \sim10$, this implies $|\hat{\xi}_{\rm V}| \sim |\hat{\xi}_{\rm U}| \sim1$ \citep[see Appendix C in][]  {bjo19}. In a homogeneous source, this leads to very high values for the circular polarisation (several tens of percent, see Appendix C), which are at least an order of magnitude larger than observed ones. 

An inhomogeneous source structure will affect the degree of circular polarisation in two different, although related, ways: (1) the initial conditions for the light rays can vary along the line of sight; at the same time, (2) these variations will also influence the effective phase difference between the characteristic waves for a given light ray. However, it is important to note that these effects are not independent but are both induced by the varying plasma properties (see Appendix C). As an example, consider a situation where the component of the magnetic field changes sign repeatedly along the line of sight (see Figure\,\ref{fig1}). This causes both the linear conversion term ($\propto K_{\rm i}K_{\rm r}$) and the phase difference between the characteristic waves ($\propto \hat{\xi}_{\rm V}$) to change sign. Although these sign-changes will lower the degree of circular polarisation of the emerging radiation, the effective value of $|\hat{\xi}_{\rm V}|$ is lowered as well.

Hence, invoking varying initial conditions to lower the observed degree of circular polarisation in the magnetic reconnection scenario discussed in \cite{pet19} is likely to lead to an effective value of $|\hat{\xi}_{\rm V}|$ substantially below unity. This, however, would be at odds with the conclusion reached above from the observed properties of the polarisation, which is most readily understood as being due to a plasma with $|\hat{\xi}_{\rm V}| \sim 10^2$. Moreover, there is another, independent argument against attributing the low observed value of the circular polarisation to large-scale cancellation. The circular polarisation varies more rapidly and with larger relative amplitude than either the flux or linear polarisation. However, it only rarely changes sign in an individual source \citep{w/d83,kom84}. To make these observations consistent with the needed large cancellation may require some fine-tuning of the source properties.

\newpage

\section{Conclusions}\label{sect6}
The interaction between a propagating electromagnetic wave and an inhomogeneous plasma can be formulated in, at least, two different ways. Normally, it is described in terms of the Stokes parameters, but an equivalent formulation can be made using the electromagnetic field itself. In the latter formulation, the concept of characteristic waves is central. The main results of the present paper are:

1) A shorter and more direct derivation of the equations describing the coupling of the characteristic waves is presented.

2) With constant coupling, these equations have a solution valid under a wide range of circumstances. This makes possible a much more simplified treatment of the effects of inhomogeneities on the emerging polarisation. In addition, the use of the polarisation properties of the characteristic waves allows a transparent formulation of the solution.

3) The effects of inhomogeneities can be substantial for nearly linear characteristic waves but rather minor for nearly circular characteristic waves. Compared to the circular polarisation from a homogeneous source, this affects mainly its magnitude and only marginally its frequency dependence. Hence, inhomogeneities have little effect on the frequency dependence of the circular polarisation.

4) The frequency dependence of the circular polarisation differs significantly for plasma properties corresponding to nearly circular and nearly linear characteristic waves. It is argued that the observed polarisation properties of compact radio sources fit nicely with nearly circular but are hard to reconcile with nearly linear characteristic waves. This, in turn, constrains the modelling of the acceleration process as well as the presence of electron-positron pairs; for example, some of the currently preferred parameter values do not easily match with observations.

\newpage

\appendix

\begin{center}
{\bf Appendix}
\end{center}

\section{Propagation of a polarised light ray in an inhomogeneous medium with a constant azimuthal angle $\phi$}
The appropriate matrix in Equation (\ref{eq19}) can be diagonalised to give the eigenvalues $\beta_{\pm} = \pm \sqrt{1+\alpha^2}$. The two characteristic waves can then be written $X_{\pm} = X_{{\rm  o},\pm} \exp(\beta_{\pm}\chi)$ and $Y_{\pm} = c_{\pm}X_{\pm}$, where $c_{\pm} = \alpha/(1+\beta_{\pm})$. With  $X = X_{+} + X_{-}$ and  $Y = Y_{+} + Y_{-}$ together with the initial values (i.e. $\chi = 0$), $X_{\rm o} = \sqrt{\frac{I_{\rm 0}}{2(1+q_{\rm o})}}(1+q_{\rm 0})$ and $Y_{\rm o} = -\sqrt{\frac{I_{\rm 0}}{2(1+q_{\rm o})}}\sigma_{\rm 0}$ \citep[see][]{bjo19}, one finds
\begin{equation}
	X_{\rm o,\pm} = \sqrt{\frac{I_{\rm 0}}{2(1+q_{\rm o})}}\times\frac{(1+q_{\rm 0} - c_{\pm}\sigma_{\rm 0})}{1+c_{\pm}^2},
	\label{A1}
\end{equation}
where $c_{\rm +}c_{\rm -} = -1$ has been used.
This leads to
\begin{eqnarray}
	X &=& \nonumber\\
	   & &\hspace{-1cm}\sqrt{\frac{I_{\rm o}}{2(1+q_{\rm o})}}\frac{1}{2\beta}[\{(\beta+1)(1+q_{\rm 0}) - \alpha \sigma_{\rm 0}\}\exp(\beta \chi) + 
	   \{(\beta -1)	(1+q_{\rm o}) + \alpha \sigma_{\rm o}\}\exp(-\beta \chi)]\nonumber \\
	Y &=& \nonumber\\
	& &\hspace{-1cm}\sqrt{\frac{I_{\rm o}}{2(1+q_{\rm o})}}\frac{1}{2\beta}[\{\alpha(1+q_{\rm 0}) + (1-\beta) \sigma_{\rm 0}\}\exp(\beta \chi) - 
	\{\alpha(1+q_{\rm o}) + (\beta + 1) \sigma_{\rm o}\}\exp(-\beta \chi)], \nonumber\\
	\label{eqA2}
\end{eqnarray}
where $\beta \equiv \sqrt{1+\alpha^2}$ has been introduced, which makes it possible to write $1 + c_{\rm \pm}^2 = 2\beta/(\beta \pm 1)$ and $c_{\rm \pm}/ (1 + c_{\rm \pm}^2) = \pm \alpha/2\beta$.  A more convenient form of Equation (\ref{eqA2}) is given by
\begin{eqnarray}
	X &=& \sqrt{\frac{I_{\rm o}}{2(1+q_{\rm o}})}\left[(1+q_{\rm o} -\alpha \sigma_{\rm o})\frac{\sinh (\beta \chi)}{\beta} + (1+q_{\rm o}) \cosh(\beta \chi)
	\right]\nonumber \\
	Y &=& \sqrt{\frac{I_{\rm o}}{2(1+q_{\rm o}})}\left[ (\alpha (1+q_{\rm o}) + \sigma_{\rm o})\frac{\sinh (\beta \chi)}{\beta} -\sigma_{\rm o} \cosh(\beta \chi)
	\right],\nonumber \\
	\label{eqA3}
\end{eqnarray}
which then leads to Equation (\ref{eq20}).

\subsection{Limiting solution for $|\beta \chi| <1$}
The relevant Stokes parameters can be obtained from Equation (\ref{eq21a}),
\begin{eqnarray}
	U&+&iV \equiv2E_{\rm x}E_{\rm y}^{*} =  \frac{I_{\rm o}}{1+q_{\rm o}}K_{\rm o}^{*}\times\nonumber\\
	& &\left[(1+q_{\rm o})\sigma_{\rm o}^{*} -
	\frac{\delta k^{*} \tau}{2}(1+q_{\rm o})^2 (1+\chi^{*})- \frac{\delta k \tau}{2}|\sigma_{\rm o}|^2 (1-\chi)\right]
	\label{eqA9}
\end{eqnarray}
The various terms in Equation (\ref{eqA9}) are most conveniently evaluated using the relations $K_{\rm o} \delta k = -(\Upsilon_{\rm V} +
i\Upsilon_{\rm L})$ and $\delta k / K_{\rm o} = (\Upsilon_{\rm V} -i \Upsilon_{\rm L})$, which can be obtained from Equations (\ref{eq3}) and (\ref{eq4}). The result is
\begin{eqnarray} 
	U+iV = I_{\rm o}\left[u_{\rm o} +iv_{\rm o} +\frac{\tau}{2}\{\Upsilon_{\rm V}^{*} -\Upsilon_{\rm V} +i(\Upsilon_{\rm L} -
	\Upsilon_{\rm L}^{*}) + q_{\rm o}(\Upsilon_{\rm V}^{*} +\Upsilon_{\rm V} -i(\Upsilon_{\rm L} +
	\Upsilon_{\rm L}^{*}))\}\right.\nonumber\\
	+\left.\frac{\tau}{2}\{\Upsilon_{\rm V}^{*}\chi^{*} + \Upsilon_{\rm V}\chi -i( \Upsilon_{\rm L}\chi +  \Upsilon_{\rm L}^{*}\chi^{*})
	+ q_{\rm o}(\Upsilon_{\rm V}^{*}\chi^{*} - \Upsilon_{\rm V}\chi + i(\Upsilon_{\rm L}\chi - \Upsilon_{\rm L}^{*}\chi^{*}))\}\right].\nonumber\\
	\label{eqA10}
\end{eqnarray}
From the definitions of $\Upsilon_{\rm V}$ and $\Upsilon_{\rm L}$, one finds
\begin{eqnarray}
	U+iV = I_{\rm o}\left[u_{\rm o} + iv_{\rm o} +\tau \{-\xi_{\rm U} + (\Upsilon_{\rm V}\chi)_{\rm r} +q_{\rm o}(\hat{\xi}_{\rm V} - 
	(\Upsilon_{\rm L}\chi)_{\rm i}\}\right.\nonumber\\
	\left. - i\tau\{\xi_{\rm V} + (\Upsilon_{\rm L}\chi)_{\rm r} +q_{\rm o}(\hat{\xi}_{\rm U} + (\Upsilon_{\rm V}\chi)_{\rm i}\}\right],
	\label{eqA11}
\end{eqnarray}
which shows that the circular polarisation is given by
\begin{equation}
	V = I_{\rm o}[v_{\rm o} -\xi_{\rm V}\tau -q_{\rm o}\hat{\xi}_{\rm U}\tau - \hat{\xi}_{\rm U}\chi_{\rm r}\tau + \xi_{\rm U}\chi_{\rm i}\tau
	-q_{\rm o}\tau(\hat{\xi}_{\rm V}\chi_{\rm i} + \xi_{\rm V}\chi_{\rm r})]
\end{equation}

\subsection{Limiting solution for $|\alpha| \gg 1$}
For $|\alpha| \gg 1$, one may expand the relevant expressions to first order in $\alpha^{-1}$. Since $\beta = \alpha$ in this limit so that $\beta \chi = \delta k\tau/2$, Equation (\ref{eqA3}) yields
\begin{eqnarray}
	2XK_{\rm o}^{*}Y^{*} &=&(2XK_{\rm o}^{*}Y^{*})_{\rm hom}
	+ \frac{I_{\rm o}K_{\rm o}^{*}}{(1+q_{\rm o})|\alpha|^2} \times\nonumber \\
	&& \left[ \{\alpha^{*}(1+q_{\rm o})\sinh (\delta k \tau/2)((1+q_{\rm o}) \sinh^{*}(\delta k\tau/2) - \sigma_{\rm o}^{*}\cosh^{*}(\delta k\tau/2))
	\} \right.\nonumber \\
	&+&\left.\{\alpha \sigma_{\rm o}^{*}\sinh^{*}(\delta k\tau/2)((1+q_{\rm o})\cosh(\delta k\tau/2) - \sigma_{\rm o}\sinh(\delta k\tau/2))\}\right].
	\nonumber \\
	\label{eqA4}
\end{eqnarray}
Here, $(2XK_{\rm o}^{*}E_{\rm y}^{*})_{\rm hom}$ is the part corresponding to a homogeneous source.
With $U+iV = -2XK_{\rm o}^{*}Y^{*}\exp (-\tau -2i\chi_{\rm i})$ (see Equation (\ref{eq18})),
\begin{eqnarray}
	U+iV &=& I_{\rm o}\exp (-\tau -2i\chi_{\rm i})\left[u_{\rm hom} + iv_{\rm hom}\right. \nonumber\\
	&+& \frac{K_{\rm o}^{*}|\sinh(\delta k\tau/2)|^2}{|\alpha|^2}\left\{\frac{\alpha(1-q_{\rm o})}{|K_{\rm o}|^2}  - \alpha^{*}(1+q_{\rm o})
	\right\}\nonumber \\
	&+& \left.\frac{2i(u_{\rm o} + iv_{\rm o})}{|\alpha|^2}\{\alpha^{*} \sinh(\delta k\tau/2) \cosh^{*}(\delta k\tau/2)\}_{\rm i}\right],\nonumber \\
	\label{eqA5}
\end{eqnarray}
where $|\sigma_{\rm o}|^2 = (1-q_{\rm o}^2)/|K_{\rm o}|^2$ has been used.

\subsection{The range of validity for the $\alpha=constant$ solution}
The two eigenfunctions corresponding to Equation (\ref{eq19}) are $X + c_{\rm \pm}Y$. The errors implied by assuming $\alpha=constant$ can be estimated by letting $c_{\rm \pm}$ vary with $\chi$. One can then write
\begin{eqnarray}
	\frac{{\rm d}(X+c_{\rm +}Y)}{{\rm d} \chi} &=& \beta(X+c_{\rm +}Y)\nonumber\\
	\frac{{\rm d}(X+c_{\rm -}Y)}{{\rm d} \chi} &=& -\beta(X+c_{\rm -}Y).\nonumber\\
	\label{eqA6}
\end{eqnarray}
With the use of the expressions for $c_{\rm \pm}$, it is found that
\begin{eqnarray}
	\frac{{\rm d}X}{{\rm d} \chi} + \frac{Y}{c_{\rm +} - c_{\rm -}}\left(\frac{{\rm d}\ln c_{\rm +}}{\rm d \chi} - \frac{{\rm d}\ln  c_{\rm -}}{{\rm d} 
	\chi}\right) &=&
	X + \alpha Y\nonumber\\
	\frac{{\rm d}Y}{{\rm d} \chi} + Y\left(\frac{{\rm d}\ln(c_{\rm +} - c_{\rm -})}{{\rm d} \chi}\right) &=&
	\alpha X - Y.\nonumber\\
	\label{eqA7}
\end{eqnarray}
This shows explicitly how the variations of $c_{\rm \pm}$ affect the propagation of a light ray. Since $c_{\rm \pm}$ are functions of $\alpha$ only, Equation (\ref{eqA7}) can be rewritten as
\begin{eqnarray}
	\frac{{\rm d}X}{{\rm d}\chi} & = & X + \alpha \left(1- \frac{1}{1+\alpha^2}\frac{{\rm d}\ln \alpha}{{\rm d}\chi}\right)Y
	\nonumber\\
	\frac{{\rm d}Y}{{\rm d}\chi} & = & \alpha X -\left(1- \frac{1}{1+\alpha^2}\frac{{\rm d}\ln \alpha}{{\rm d}\chi}\right)Y.
	\label{eqA8}
\end{eqnarray}
Comparison to Equation (\ref{eq19}) makes it clear that so long as $|(1+\alpha^2)^{-1}({\rm d} \ln \alpha / {\rm d} \chi)| \ll 1$, the approximation $\alpha = constant$ is expected to be a good one.

\newpage

\section{Propagation of a polarised light ray in a medium with a varying azimuthal angle $\phi$}
The transport equation in this case can be solved following the same procedure as in Appendix A. Diagonalising the appropriate matrix  for Equation (\ref{eq35}) gives eigenvalues $\beta_{\rm \pm} = \pm\sqrt{\alpha^2 -1}$. Likewise, the corresponding relation between $X_{\rm \pm}$ and $Y_{\rm \pm}$ is given by $c_{\rm \pm} = \beta_{\rm \pm}/(\alpha -1)$. Since the initial values are the same, this leads to
\begin{equation}
	X_{\rm o,\pm} = \sqrt{\frac{I_{\rm o}}{2(1+q_{\rm o})}}\times \frac{1}{2}\left(1+q_{\rm o} - \frac{\sigma_{\rm o}}{c_{\rm \pm}}\right).
	\label{eqB1}
\end{equation}
With $\beta \equiv \sqrt{\alpha^2 -1}$, the solution to Equation (\ref{eq35}) can be written
\begin{eqnarray}
	X &=& \sqrt{\frac{I_{\rm o}}{2(1+q_{\rm o})}}\frac{1}{2}\left[\{1+q_{\rm o} - \frac{(\alpha - 1)\sigma_{\rm o}}{\beta}\}\exp(\beta \chi) +
	\{1+q_{\rm o} + \frac{(\alpha - 1)\sigma_{\rm o}}{\beta}\}\exp(-\beta \chi)\right]\nonumber\\
	Y &=& \sqrt{\frac{I_{\rm o}}{2(1+q_{\rm o})}}\frac{1}{2}\left[\{(1+q_{\rm o})\frac{(\alpha +1)}{\beta} - \sigma_{\rm o}\}\exp(\beta \chi) -
	\{(1+q_{\rm o})\frac{(\alpha +1)}{\beta} + \sigma_{\rm o}\exp(-\beta \chi)\}\right].\nonumber\\
	\label{eqB2}
\end{eqnarray}
This can be rewritten as
\begin{eqnarray}
	X &=& \sqrt{\frac{I_{\rm o}}{2(1+q_{\rm o})}}\left[(1+q_{\rm o})\cosh(\beta\chi) + (1-\alpha )\sigma_{\rm o}\frac{\sinh(\beta\chi)}
	{\beta}\right]\nonumber\\
	Y &=& \sqrt{\frac{I_{\rm o}}{2(1+q_{\rm o})}}\left[-\sigma_{\rm o}\cosh(\beta\chi) + (1+\alpha )(1+q_{\rm o})\frac{\sinh(\beta\chi)}
	{\beta}\right].\nonumber\\
	\label{eqB3}
\end{eqnarray}
The corresponding electric field (i.e., Equation \ref{eq36}) is then obtained by inserting these expressions into Equation (\ref{eq33}).

\subsection{Limiting solution for $|\alpha| \gg 1$}
For $|\alpha| \gg 1$, the electric field in Equation (\ref{eq36}) may be expanded to first order in $\alpha^{-1}$. Similar to the $\varphi = constant$ case, in this limit, $\beta = \alpha$ and $\beta\chi = \delta k \tau/2$. Keeping first order terms in  $\alpha^{-1}$ and $\chi$, this yields
\begin{eqnarray}
	U+iV &\equiv& 2E_{\rm x}E_{\rm y}^{*} = \frac{I_{\rm o}}{1+q_{\rm o}}\exp(-\tau)K_{\rm o}^{*}\times[u_{\rm hom} + iv_{\rm hom} +
	\nonumber\\
	&+&\left\{\frac{\alpha^{*}}{|\alpha|^2}\sigma_{\rm o}\sinh(\delta k\tau/2) + \chi(-\sigma_{\rm o}\cosh(\delta k\tau/2) + 
	(1+q_{\rm o})\sinh(\delta k\tau/2)\right\}\nonumber\\
	&&\times\left\{\sigma_{\rm o}^{*}\cosh^{*}(\delta k\tau/2) - (1+q_{\rm o})\sinh^{*}(\delta k\tau/2)\right\}\nonumber\\
	&+&\left\{-\frac{\alpha}{|\alpha|^2}(1+q_{\rm o})\sinh^{*}(\delta k\tau/2) + \chi^{*}((1+q_{\rm o})\cosh^{*}(\delta k\tau/2) 
	-\sigma^{*}_{\rm o}\sinh^{*}(\delta k\tau/2))\right\},\nonumber\\
	&&\left.\times\left\{(1+q_{\rm o})\cosh(\delta k\tau/2) - \sigma_{\rm o}\sinh(\delta k\tau/2)\right\}\right]\nonumber\\
	\label{eqB4}
\end{eqnarray}
where, again, the subscript "hom" refers to the corresponding homogeneous term. This can be rewritten as
\begin{eqnarray}
	U+iV &=& I_{\rm o}\exp(-\tau)K_{\rm o}^{*}\times[u_{\rm hom} + iv_{\rm hom} +
	\frac{(\alpha - \alpha^{*})}{|\alpha|^2}\sigma_{\rm o}|\sinh(\delta k \tau/2)|^2\nonumber\\
	&-&\frac{\alpha}{|\alpha|^2}(1+q_{\rm o})\cosh(\delta k \tau/2)\sinh^{*}(\delta k \tau/2) + \frac{\alpha^{*}}{|\alpha|^2}
	\frac{(1-q_{\rm o})}{|K_{\rm o}|^2}\cosh^{*}(\delta k \tau/2)\sinh(\delta k \tau/2)\nonumber\\
	&-&\chi\left\{\frac{(1-q_{\rm o})}{|K_{\rm o}|^2}|\cosh(\delta k \tau/2)|^2+ (1+q_{\rm o})|\sinh(\delta k \tau/2)|^2 \right\}\nonumber\\
	&+&\chi^{*}\left\{(1+q_{\rm o})|\cosh(\delta k \tau/2)|^2+ \frac{(1-q_{\rm o})}{|K_{\rm o}|^2}|\sinh(\delta k \tau/2)|^2 \right\}\nonumber\\
	&+& \chi\left\{\sigma_{\rm o}\cosh(\delta k \tau/2)\sinh^{*}(\delta k \tau/2) + \sigma_{\rm o}^{*}\cosh^{*}(\delta k \tau/2)\sinh(\delta k \tau/ 
	2)\right\}\nonumber\\
	&-& \chi^{*}\left\{\sigma_{\rm o}^{*}\cosh(\delta k \tau/2)\sinh^{*}(\delta k \tau/2) + \sigma_{\rm o}\cosh^{*}(\delta k \tau/2)
	\sinh(\delta k \tau/ 2)\right\}\nonumber\\
	\label{eqB5}
\end{eqnarray}
In order to emphasise the role played by the initial conditions, the various terms in Equation (\ref{eqB5}) can be rearranged as follows:
\begin{eqnarray}
	U + iV &=& I_{\rm o} \exp(-\tau)[u_{\rm hom} + iv_{\rm hom} + 2i(u_{\rm o} + iv_{\rm o})\left\{\chi \cosh^{*}(\delta k \tau/2)
	\sinh(\delta k \tau/2)\right\}_{\rm i}\nonumber\\
	&+&2i\sigma_{\rm o} K_{\rm o}^{*}\left\{\frac{\alpha}{|\alpha|^2}|\sinh(\delta k \tau/2)|^2 + \chi \cosh(\delta k \tau/2)
	\sinh^{*}(\delta k \tau/2)\right\}_{\rm i}\nonumber\\
	&-&2iK_{\rm o}^{*}\left\{\chi_{\rm i} \cosh(\delta k_{\rm r} \tau) + \frac{\{\alpha \cosh(\delta k \tau/2)\sinh^{*}
	(\delta k \tau/2)\}_{\rm i}}{|\alpha|^2}\right\}\nonumber\\
	&+&2q_{\rm o}K_{\rm o}^{*}\left\{\chi_{\rm r} \cos(\delta k_{\rm i} \tau) - \frac{\{\alpha \cosh(\delta k \tau/2)\sinh^{*}
	(\delta k \tau/2)\}_{\rm r}}{|\alpha|^2}\right\}\nonumber\\
	&+&\frac{K_{\rm o}^{*}(1-|K_{\rm o}|^2)}{|K_{\rm o}|^2}(1-q_{\rm o})\times\nonumber\\
	&& \left.\left\{\frac{\alpha^{*}}{|\alpha|^2}\cosh^{*}(\delta k\tau/2)
	\sinh(\delta k\tau/2) - \chi|\cosh(\delta k\tau/2)|^2 + \chi^{*}|\sinh(\delta k\tau/2)|^2\right\}\right].\nonumber\\
	\label{eqB6}
\end{eqnarray}
It is seen that in the limit $|\delta k|\tau \ll 1$, the last four inhomogeneous terms in Equation (\ref{eqB6}) vanish to order $|\delta k\tau \chi|$. Hence, only the first inhomogeneous term contributes to the polarisation in this limit (see Equation \ref{eq38}). Furthermore, for nearly linear or nearly circular characteristic waves, $|1-|K_{\rm o}|^2| \ll |K_{\rm o}|^2$ so that the last inhomogeneous term can be neglected also for $|\delta k|\tau/2 \gsim 1$; i.e., this term will never contribute significantly to the polarisation.

\newpage
\section{Relating the plasma properties to the polarisation of the characteristic waves}
As mentioned in the main text, the homogeneous solution is often also quite useful for inhomogeneous sources. Focussing on transport induced effects (i.e., setting $v_{\rm o} = 0$) and a synchrotron plasma ($q_{\rm o} = 0$), the circular polarisation for a given light ray can be written \citep{bjo19},
\begin{eqnarray}
\lefteqn{V =  I_{\rm o} \exp(-\kappa s)\left[- u_{\rm o}\frac{K_{\rm i}K_{\rm r}}{|K|^2}\left\{\cosh(\delta k_{\rm r}\tau) - \cos(\delta k_{\rm i}
	\tau)\right\}\right.} \hspace{2.4cm}
	\nonumber\\
	& &+ \frac{K_{\rm i}}{2}\left\{\frac{|K|^2 + 1}{|K|^2} \right\}\sinh(\delta k_{\rm r}\tau) \nonumber\\
	& &+ \left.\frac{K_{\rm r}}{2}\left\{\frac{|K|^2 - 1}{|K|^2} \right\}\sin(\delta k_{\rm i}\tau) \right],
	\label{eqc1}
\end{eqnarray} 
where, for an inhomogeneous source,  $\delta k \tau = \int_0^s \Delta k {\rm d}\hat{s}$ (see Section\,\ref{sect3a}), and $K$ is the initial value (i.e., at $s=0$) of the polarisation for the characteristic waves.

With
\begin{eqnarray}
K^2 &=& \frac{1-\rho}{1+\rho}\nonumber\\
	&=&\frac{1 - |\rho|^2 -2i\rho_{\rm i}}{1 + |\rho|^2 + 2\rho_{\rm r}}
	\label{eqc2}
\end{eqnarray}	
and $K^2 = K_{\rm r}^2 - K_{\rm i}^2 + 2iK_{\rm r}K_{\rm i}$, one identifies
\begin{equation}
K_{\rm r}^2 - K_{\rm i}^2 = \frac{1 - |\rho|^2}{1 + |\rho|^2 + 2\rho_{\rm r}} \hspace{1cm} {\rm and} \hspace{1cm} K_{\rm r}K_{\rm i} = 
	\frac{-\rho_{\rm i}}{1 + |\rho|^2 + 2\rho_{\rm r}}.
	\label{eqc3}
\end{equation}
Likewise, one finds
\begin{equation}
|K|^2 = \sqrt{\frac{1 + |\rho|^2 - 2\rho_{\rm r}}{1 + |\rho|^2 + 2\rho_{\rm r}}},
\label{eqc4}
\end{equation}
so that
\begin{equation}
\frac{K_{\rm r}^2 - K_{\rm i}^2}{|K|^2} = \frac{1 - |\rho|^2}{\sqrt{1 + |\rho|^2)^2 - 4\rho_{\rm r}}} \hspace{1cm} {\rm and} \hspace{1cm} 
	\frac{K_{\rm r}K_{\rm i}}{|K|^2} = \frac{-\rho_{\rm i}}{\sqrt{1 + |\rho|^2)^2 -4\rho_{\rm r}}}.
	\label{eqc5}
\end{equation}

The first term in Equation (\ref{eqc1}) accounts for the conversion of linear to circular polarisation; its magnitude is determined by $\rho_{\rm i}$ (Equation \ref{eqc5}). The last term in  Equation (\ref{eqc1}) is zero for $|K| = 1$, which requires $\rho_{\rm r} = 0$ (Equation \ref{eqc4}). This corresponds to orthogonal characteristic waves. Furthermore, the maximum value of $|K_{\rm r}K_{\rm i}|/|K|^2$ occurs for $|K_{\rm r,i}/K_{\rm i,r}| = 1/ \sqrt{2}$, which leads to  $|K_{\rm r}K_{\rm i}|/|K|^2 = \sqrt{2}/3$. This is close to where $|K_{\rm r}| = |K_{\rm i}|$, i.e., $|\rho| = 1$. Hence, the conversion of linear to circular polarisation attains a maximum in the region where $|\rho| \approx 1$ and can reach several tens of percent. 

For a synchrotron plasma, $\xi_{\rm U} \approx 1$, while $|\xi_{\rm V}| \ll 1$. In order to be consistent with $v_{\rm o} = 0$, the circular absorptivity should be set to zero, i.e., $|\xi_{\rm V}| = 0$. The plasma properties are then described by
\begin{equation}
	\rho = \frac{\hat{\xi}_{\rm V}(\xi_{\rm U} +i\hat{\xi}_{\rm U})}{\xi_{\rm U}^2 + \hat{\xi}_{\rm U}^2}.
	\label{eqc6}
\end{equation}
This leads to
\begin{equation}
	|\rho|^2 = \frac{\hat{\xi}_{\rm V}^2}{\xi_{\rm U}^2 + \hat{\xi}_{\rm U}^2}, \hspace{1.5cm} \frac{\rho_{\rm r}}{|\rho|^2} = 
	\frac{\xi_{\rm U}}{\hat{\xi}_{\rm V}} \hspace{0.5cm} {\rm and} \hspace {0.5cm} \frac{\rho_{\rm i}}{|\rho|^2} = 
	\frac{\hat{\xi}_{\rm U}}{\hat{\xi}_{\rm V}}.
	\label{eqc7}
\end{equation}
Furthermore, it is convenient to also express the phase difference between the characteristic waves in terms of $\rho$,
\begin{equation}
	\frac{\Delta k}{\kappa} =i \hat{\xi}_{\rm V}\sqrt{1 + \frac{(\rho_{\rm i}^2 - \rho_{\rm r}^2 + 2i\rho_{\rm i}\rho_{\rm r})}
	{|\rho|^4}}
	\label{eqc8}
\end{equation}
It is seen from Equation (\ref{eqc7}) that $\rho_{\rm r}$ is a measure of the linear absorption. An important point to note from Equation (\ref{eqc8}) is that the  $\rho_{\rm r}$-dependence of $\delta k_{\rm r}$ implies $\delta k_{\rm r} = 0$ when $\rho_{\rm r} = 0$. Hence, neglect of absorption or, equivalently, assuming orthogonal characteristic waves causes the last two terms in Equation (\ref{eqc1}) to become zero. 

The contributions to the circular polarisation in a synchrotron plasma from the various terms in Equation (\ref{eqc1}) for $|\rho| \ll 1$ and $|\rho| \gg 1$ have been discussed in \cite{bjo19}. When $|\rho| \sim 1$, Equation (\ref{eqc7}) implies $|\hat{\xi}_{\rm V}| \sim |\hat{\xi}_{\rm U}|\,\gsim\,\xi_{\rm U}$. Since $\rho_{\rm r}/\rho_{\rm i} = \xi_{\rm U}/\hat{\xi}_{\rm U}$, the above discussion shows that for $\xi_{\rm U}/|\hat{\xi}_{\rm U}|\ll 1$, the main contribution comes from linear conversion. Only when $\xi_{\rm U}/|\hat{\xi}_{\rm U}| \sim 1$ do all three terms contribute substantially. Actually, this latter case may be the relevant one for compact radio sources, since $|\hat{\xi}_{\rm U}|\sim 1$ is expected for a rather large range of plasma properties. This is due to the fact that, in contrast to $\hat{\xi}_{\rm V}$, the value of $\hat{\xi}_{\rm U}$ is rather insensitive to variations in the synchrotron plasma. Moreover, this suggests that the polarisation of the characteristic waves is determined mainly by the value of $\hat{\xi}_{\rm V}$.

\clearpage

\end{document}